\documentclass{aa}

\usepackage{graphicx}
\usepackage{lscape}
\usepackage{txfonts}
\usepackage{xcolor} 
\usepackage[colorlinks=true, citecolor=blue, linkcolor=blue, urlcolor=blue]{hyperref}
\usepackage{amsmath}
\begin{document}

   \title{TOI-1135\,b: A young hot Saturn-size planet orbiting a solar-type star}


   \author{
 M.\,Mallorqu\'in\inst{\ref{i:iac},\ref{i:ull}}, 
 N.\,Lodieu\inst{\ref{i:iac},\ref{i:ull}},  
 V.\,J.\,S.\,B\'ejar\inst{\ref{i:iac},\ref{i:ull}}, 
 M.\,R.\,Zapatero Osorio\inst{\ref{i:cab}},
 J.\,Sanz-Forcada\inst{\ref{i:cab}}, 
 M.\,R.\,Alarcon\inst{\ref{i:iac},\ref{i:ull}}, 
 H.\,M.\,Tabernero\inst{\ref{i:ucm}}, 
 E.\,Nagel\inst{\ref{i:got}}, 
 K.\,A.\,Collins\inst{\ref{i:cam}}, 
 D.\,R.\,Ciardi\inst{\ref{i:cal}}, 
 M.\,Serra-Ricart\inst{\ref{i:lb},\ref{i:iac},\ref{i:ull}}, 
 J.\,Orell-Miquel\inst{\ref{i:iac}, \ref{i:ull}}, 
 K.\,Barkaoui\inst{\ref{i:uli}, \ref{i:mit}, \ref{i:iac}}, 
 A.\,Burdanov\inst{\ref{i:mit}}, 
 J.\,de\,Wit\inst{\ref{i:mit}}, 
 M.\,E.\,Everett\inst{\ref{i:tuc}}, 
 M.\,Gillon\inst{\ref{i:uli}}, 
 E.\,L.\,N.\,Jensen\inst{\ref{i:swar}}, 
 L.\,G.\,Murphy\inst{\ref{i:kut}}, 
 P.\,A.\,Reed\inst{\ref{i:kut}}, 
 B.\,Safonov\inst{\ref{i:sai}}, 
 I.\,A.\,Strakhov\inst{\ref{i:sai}}, 
 C.\,Ziegler\inst{\ref{i:naco}} 
            }
   \authorrunning{Mallorqu\'in, M., et al.}   
  \institute{
        \label{i:iac} Instituto de Astrof\'isica de Canarias (IAC), Calle V\'ia L\'actea s/n, 38205 La Laguna, Tenerife, Spain,
             \email{mmd@iac.es}
  	\and 
        \label{i:ull} Departamento de Astrof\'isica, Universidad de La Laguna (ULL), 38206 La Laguna, Tenerife, Spain
        \and
        \label{i:cab} Centro de Astrobiolog\'ia (CAB), CSIC-INTA, ESAC Campus, Camino bajo del castillo s/n, 28692, Villanueva de la Cañada, Madrid, Spain
  	\and 
        \label{i:ucm} Departamento de F\'isica de la Tierra y Astrof\'isica and IPARCOS-UCM (Instituto de F\'isica de Part\'iculas y del Cosmos de la UCM), Facultad de Ciencias F\'isicas, Universidad Complutense de Madrid, 28040, Madrid, Spain
        \and
        \label{i:got} Institut f\"ur Astrophysik und Geophysik, Georg-August-Universit\"at G\"ottingen, Friedrich-Hund-Platz 1, 37077 G\"ottingen, Germany
   	\and 
        \label{i:cam} Center for Astrophysics, Harvard \& Smithsonian, 60 Garden Street, Cambridge, MA 02138, USA
  	\and 
        \label{i:cal} NASA Exoplanet Science Institute - Caltech/IPAC, Pasadena, CA USA
        \and
        \label{i:lb} Light Bridges S.\,L.\,, Avda. Alcalde Ram\'irez Bethencourt 17, 35004 Las Palmas de Gran Canaria, Canarias, Spain
  	\and 
        \label{i:uli} Astrobiology Research Unit, Universit\'e de Li\`ege, 19C All\'ee du 6 Ao\^ut, 4000 Li\`ege, Belgium
  	\and 
        \label{i:mit} Department of Earth, Atmospheric and Planetary Science, Massachusetts Institute of Technology, 77 Massachusetts Avenue, Cambridge, MA 02139, USA
  	\and 
        \label{i:tuc} NSF’s National Optical-Infrared Astronomy Research Laboratory, 950 N. Cherry Ave., Tucson, AZ 85719, USA
  	\and 
        \label{i:swar} Department of Physics \& Astronomy, Swarthmore College, Swarthmore PA 19081, USA
  	\and 
        \label{i:kut} Department of Physical Sciences, Kutztown University, Kutztown PA 19530, USA
  	\and 
        \label{i:sai} Sternberg Astronomical Institute Lomonosov Moscow State University
  	\and 
        \label{i:naco} Department of Physics, Engineering and Astronomy, Stephen F. Austin State University, 1936 North St, Nacogdoches, TX 75962, USA
       }
  
  \date{Received 19 December 2023 / Accepted 15 February 2024}

  \abstract
   {
   Despite the thousands of planets in orbit around stars known to date, the mechanisms of planetary formation, migration, and atmospheric loss remain unresolved. In this work, we confirm the planetary nature of a young Saturn-size planet transiting a solar-type star every 8.03 d, TOI-1135\,b. The age of the parent star is estimated to be in the interval of 125--1000 Myr based on various activity and age indicators, including its stellar rotation period of 5.13\,$\pm$\,0.27 d and the intensity of photospheric lithium. We obtained follow-up photometry and spectroscopy, including precise radial velocity measurements using the CARMENES spectrograph, which together with the TESS data allowed us to fully characterise the parent star and its planet. As expected for its youth, the star is rather active and shows strong photometric and spectroscopic variability correlating with its rotation period. We modelled the stellar variability using Gaussian process regression. We measured the planetary radius at 9.02\,$\pm$\,0.23 R$_\oplus$ (0.81\,$\pm$\,0.02 R$_{\mathrm{Jup}}$) and determined a 3$\sigma$ upper limit of $<$\,51.4 M$_\oplus$ ($<$\,0.16 \,M$_{\rm{Jup}}$) on the planetary mass by adopting a circular orbit. Our results indicate that TOI-1135\,b is an inflated planet less massive than Saturn or Jupiter but with a similar radius, which could be in the process of losing its atmosphere by photoevaporation. This new young planet occupies a region of the mass-radius diagram where older planets are scarse, and it could be very helpful to understanding the lower frequency of planets with sizes between Neptune and Saturn.
   }

  \keywords
   {
    planetary systems -- planets and satellites: individual: TOI-1135\,b -- planets and satellites: gaseous planets -- methods: radial velocity -- techniques: spectroscopic -- stars: solar-type
   }
   \maketitle
%

\section{Introduction}
\label{sec:intro}

Close gas giants represent approximately 10\% of the total population of known exoplanets.\footnote{\url{http://exoplanet.eu/}} However, how short-period gas giants form remains an open question. Although the family of gas giants is large, their occurrence rate is relatively low, with $\lesssim$1\% orbiting a solar-type star and even less for later stars \citep{Wright2012}. This contradiction is explained because they are the easiest exoplanets to detect by both transit and radial velocity (RV) methods. 

Different formation mechanisms for these planets have been proposed \citep{Dawson2018}, including in situ formation \citep{Batygin2016}; disc migration \citep{Lin1996}; and high eccentricity tidal migration. The migration theory provides a reasonable explanation for the existence of short-orbit gas giants. According to this theory, these planets are believed to form farther out from their host stars (beyond the snowline), where the protoplanetary discs are rich in gas and dust. Through various mechanisms, such as gravitational interactions with other planets or interactions with the gas disc, these planets undergo a process of inwards migration towards the star until ending up in close-in orbits. However, the details of the migration mechanisms and the precise conditions required for their formation and subsequent migration are still being actively studied. Continued observations are crucial for unravelling the origin, formation, and evolution processes of these uncommon exoplanets.

To understand the existence of close gas giant planets, we must constrain their formation and evolution models. To do so, it is necessary to find young exoplanets that provide observational support of these models and constraining of their timescales. However, such stars present a high level of stellar activity, which is an obstacle to accurately characterising and measuring the properties of young exoplanets. The photometric and RV variations caused by this intense stellar activity are several times larger than the Keplerian signals attributed to the planets themselves \citep{Trevor2019, plav20, cale21, Suarez-Mascareno2022}. Therefore, the activity-induced signals can overshadow or mimic the planetary signals, making it challenging to separate and extract the true planetary properties from the observed data. It requires careful analysis and sophisticated techniques to disentangle the planetary signals from the noise and/or systematics and activity-induced variations \citep{VonStorch1999, Rajpaul2015, Barros2020, Perger2021}.

In recent years, about twenty giant planets (with sizes larger than Neptune) have been found orbiting stellar members of Tuc-Hor \citep{Newton2019}, $\beta$\,Pic \citep{plav20, Martioli2021}, and Pisces-Eridanus \citep{newton21}; young moving groups IC 2602 \citep{Bouma2020} and $\delta$\;Lyr \citep{bouma22a} open clusters; the Tau-Aur region \citep{Trevor2019}; the Sco-Cen OB association \citep{Rizzuto2020} and its sub-components Upper Scorpius \citep{Mann2016} and Lower Centaurus Crux \citep{Mann2022}; and other young field stars \citep{Sanchis-Ojeda2013, sun19, zhou21, Subjak2022, Kabath2022, dai23, Heitzmann2023} targeted by the Kepler \citep{Kepler}, K2 \citep{K2}, and the Transiting Exoplanet Survey Satellite \citep[TESS; ][]{TESS} space missions. Of the aforementioned young planetary systems, only four have mass measurements obtained through RV observation campaigns and, consequently, densities: AU\,Mic\,b \citep[$\sim$20\,Myr;][]{klein21, cale21, zicher22, klein22, Donati2023}; V1298\,Tau\,b and e \citep[$\sim$20 Myr;][]{Suarez-Mascareno2022, Sikora2023, Finociety2023}; TOI-1268\,b \citep[100--380 Myr;][]{Subjak2022}; and TOI-4562\,b \citep[$<$700 Myr;][]{Heitzmann2023}, where TOI-1268\,b is the only hot Jupiter (P$_{\rm{orb}}$\,$=$\,8.16\,days, M$_{\rm{p}}$\,$=$\,0.29\,M$_{\rm{Jup}}$) in the sample.

An alert was released by the TESS Science Office (TSO) on 27 August 2019 regarding the transit signal observed in TOI-1135\@. This transit signal was identified with a period of 8.0277 days and a depth of 6.347\,$\pm$\,0.274 parts per thousand (ppt). The TOI-1135.01 planet candidate has a radius of 9.46\,$\pm$\,0.48 times the radius of Earth (R$_{\oplus}$) and an equilibrium temperature of 1459\,K. At the time of writing, there have been no alerts indicating the presence of another planetary candidate in the system. During the preparation of this work, \cite{Hord2023} proposed a list of the `best-in-class' TOIs for atmospheric characterisation with the James Webb Space telescope (JWST). In their work, they vetted and statistically validated dozens of candidates, classifying TOI-1135\,b as a statistically validated planet.

This paper introduces the discovery and mass characterisation of a close giant gas planet orbiting the young solar-type star TOI-1135, with an orbital period of 8.03 days. The paper is structured as follows. In Sects.\ \ref{sec:tesslc} and \ref{sec:Observations}, we provide a description of the TESS photometry, and we present ground-based follow-up observations of the system. In Sect.\ \ref{sec:properties}, we focus on determining the physical properties of the star. We perform photometric, RV, and transit time variation (TTV) analysis in Sect.\ \ref{sec:analysis}. False positive scenarios are discussed in Sect.\ \ref{sec:fap}. In Sect.\ \ref{sec:disc}, we discuss the composition of the planet and its key implications. Finally, in Sect.\ \ref{sec:concl}, we summarize the main results of our study. 

\section{TESS photometry}
\label{sec:tesslc}

TOI-1135 was observed by TESS \citep{TESS} in eight sectors, each one in a two-minute short cadence integration. Specifically, it was observed in sectors 14, 19, 20, and 26 during the TESS primary mission; in sectors 40, 47, and 53 during the first extended mission; and in sector 59 during the second extended mission. Additionally, TESS is scheduled to re-observe TOI-1135 in sectors 74 and 79, and this will take place in 2024\@.

\begin{figure}[ht!]
\includegraphics[width=1\linewidth]{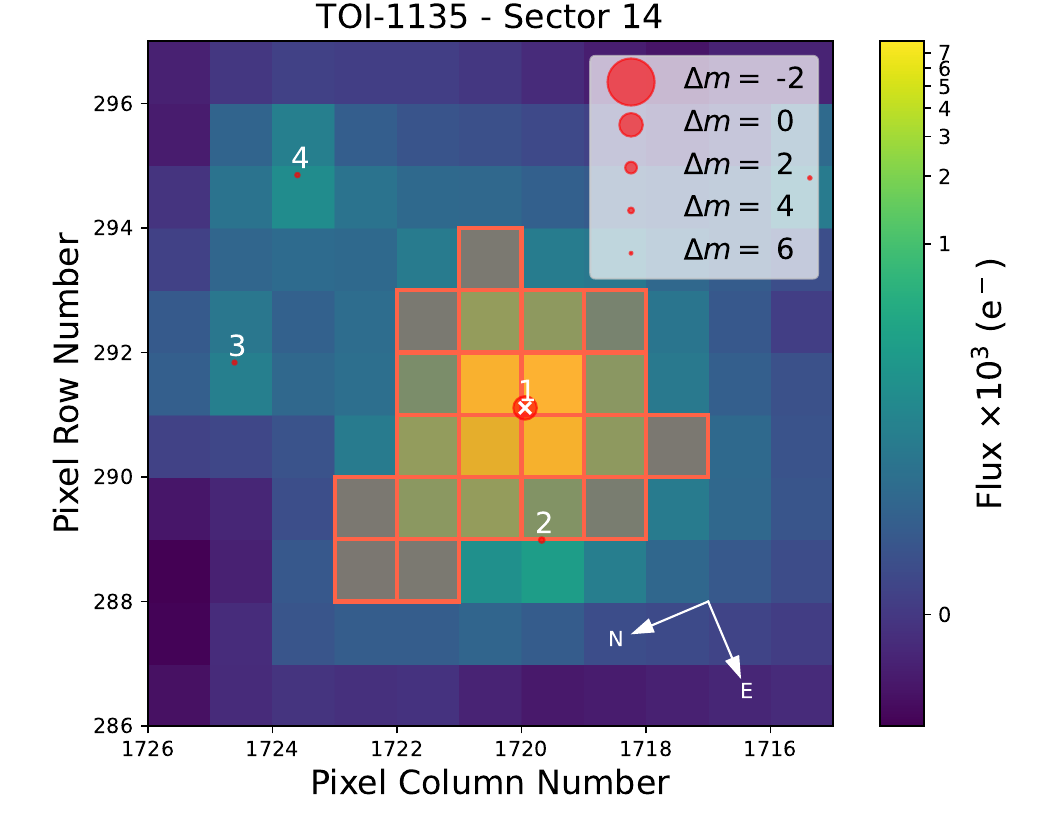}
\caption{Target pixel file plot for TOI-1135 in sector 14\@. The red squares indicate the optimal photometric aperture that was selected to obtain the SAP flux. The TESS pixel scale is 21\arcsec. Additionally, the $G$-band magnitudes from $Gaia$ DR3 are represented as red circles of different sizes, highlighting stars close to TOI-1135 up to 6 magnitudes fainter.
\label{fig:TPF}}
\end{figure}

\begin{figure*}[ht!]
\includegraphics[width=1\linewidth]{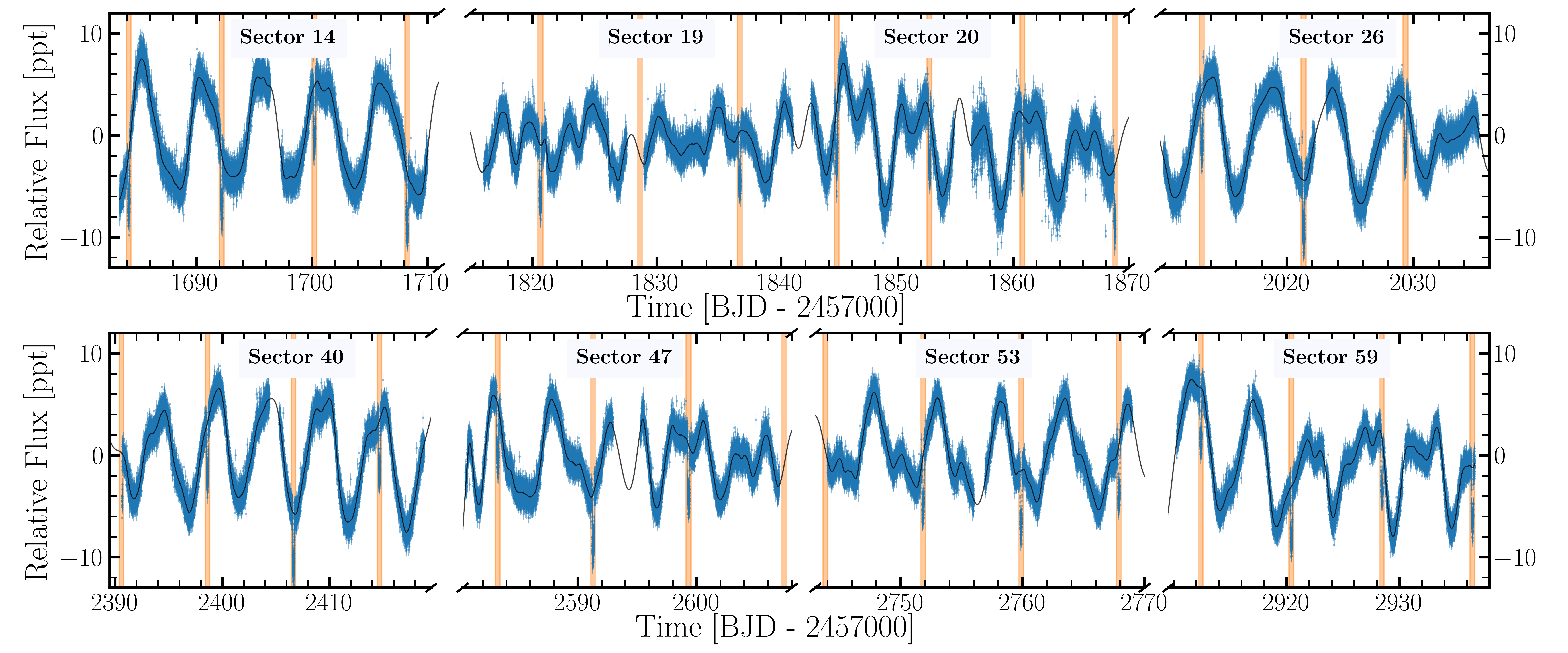}
\caption{Light curves of TOI-1135 for the eight sectors of TESS. The PDCSAP flux is represented by the blue dots, while the black line represents the best-fit model. Additionally, vertical orange lines indicate the timing of planetary transits for TOI-1135\,b.
\label{fig:LC_TESS}}
\end{figure*}

All sectors were processed by the Science Processing Operations Center (SPOC; \citealp{SPOC}) photometry and transit search pipeline at the NASA Ames Research Center. The light curves and target pixel files (TPFs) were downloaded from the Mikulski Archive for Space Telescopes\footnote{\url{https://archive.stsci.edu/}} (MAST), which provides both the simple aperture photometry (SAP) and the pre-search data conditioning SAP flux (PDCSAP). We plotted in Fig.\ \ref{fig:TPF} the TPF of sector 14 using the \texttt{tpfplotter}\footnote{\url{https://github.com/jlillo/tpfplotter}} \citep{aller2020} tool. Overlaid on the TPF is the $Gaia$ Data Release 3 (DR3) catalogue \citep{gaia2016, gaiadr3}, which includes bright sources down to 6 mag fainter than our target. We searched for potential sources of contamination and confirmed that source number 2, located at $\sim$40\arcsec{} and with a $\Delta$m$\sim$4 mag, contaminates the selected photometric aperture. However, ground-based observations show that TOI-1135 is the source of the intense stellar activity and of the transits observed by TESS (Sect.\ \ref{sec:fap}). For the rest of our analysis, we used the PDCSAP flux, which was corrected for instrumental errors and crowding. Nevertheless, we performed tests and found that our analysis and results using the SAP flux light curves are comparable and that the PDCSAP flux does not remove or change the stellar activity signal due to rotation. Figure\ \ref{fig:LC_TESS} illustrates the PDCSAP flux light curve of TOI-1135 for all TESS sectors along with the best-fit model (see Sect.\ \ref{sec:joint} for details). The TESS data have a dispersion of $\sigma_{\mathrm{TESS}}$\,$\sim$\,3.5 ppt, whereas the average error bar is $\sim$0.7 ppt. There is remarkable periodical variability with varying amplitudes, and the maximum peak-to-peak variation is $\sim$15 ppt. No apparent flares are evident in the data.

\section{Ground-based follow-up observations}
\label{sec:Observations}

\subsection{NESSI speckle imaging}
\label{sec:nessi}

We used the speckle imager NESSI \citep{Scott2018} at the WIYN 3.5\,m telescope to obtain high-resolution images of TOI-1135 on 21 April 2022 as part of the speckle imaging queue at WIYN\@. NESSI is normally operated as a dual-channel imager, but it was reduced to one camera during this observing run due to an optical alignment problem. We observed in the NESSI red channel using a 40\,nm wide filter with a central wavelength of 832\,nm. The target was centred in a 256$\times$256\,pixel (4.6\arcsec$\times$4.6\arcsec) subregion of the CCD and observed in a series of 9000 frames of 40\,ms at a plate scale near 0.02\arcsec\,pixel$^{-1}$. The TOI-1135 observations were preceded by a 1000-frame observation of HD\,112014 to serve as a point source calibrator. Additionally, during the queue run, various binary stars with well-established astrometric properties were observed to calibrate the instrument's plate scales and orientations.

\begin{figure}[ht!]
\includegraphics[width=1\linewidth]{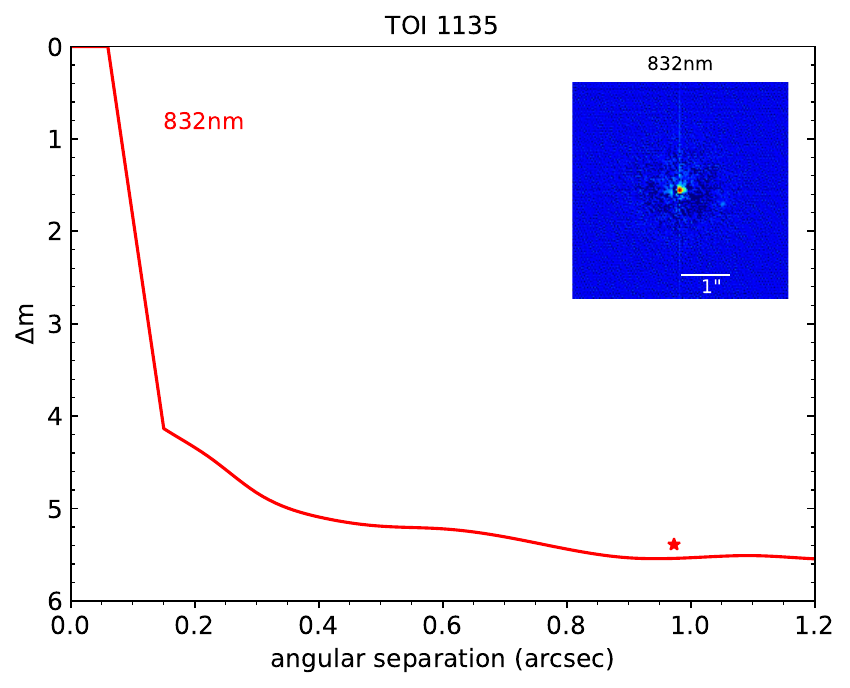}
\caption{Contrast curve of TOI-1135 from the NESSI instrument in the 832\,nm band. The inset figure shows the reconstructed 2.3$\times$2.3\arcsec{}$^2$. A fainter companion was observed at 0.97\arcsec (red star).
\label{fig:NESSI}}
\end{figure}

The NESSI speckle data were reduced using a custom speckle data reduction pipeline described by \cite{Howell2011}. This pipeline produces high-level data products, including reconstructed images of the field around the target and contrast curves measured from those images. Other data products include astrometric measurements of secondary sources detected in proximity to the target. The astrometric and photometric properties of secondary sources are measured from the power spectrum of the data, or Fourier transform of the mean of the auto-correlation functions of each speckle frame. In the case of TOI-1135, a companion 5.39 mag fainter than the source was detected at a separation of 0.973\arcsec (Fig.\ \ref{fig:NESSI}). The position angle of the companion, measured north through east relative to TOI-1135, was 252.584\,degrees.

\subsection{SAI speckle imaging}
\label{sec:sai}

\begin{figure}[ht!]
\includegraphics[width=1\linewidth]{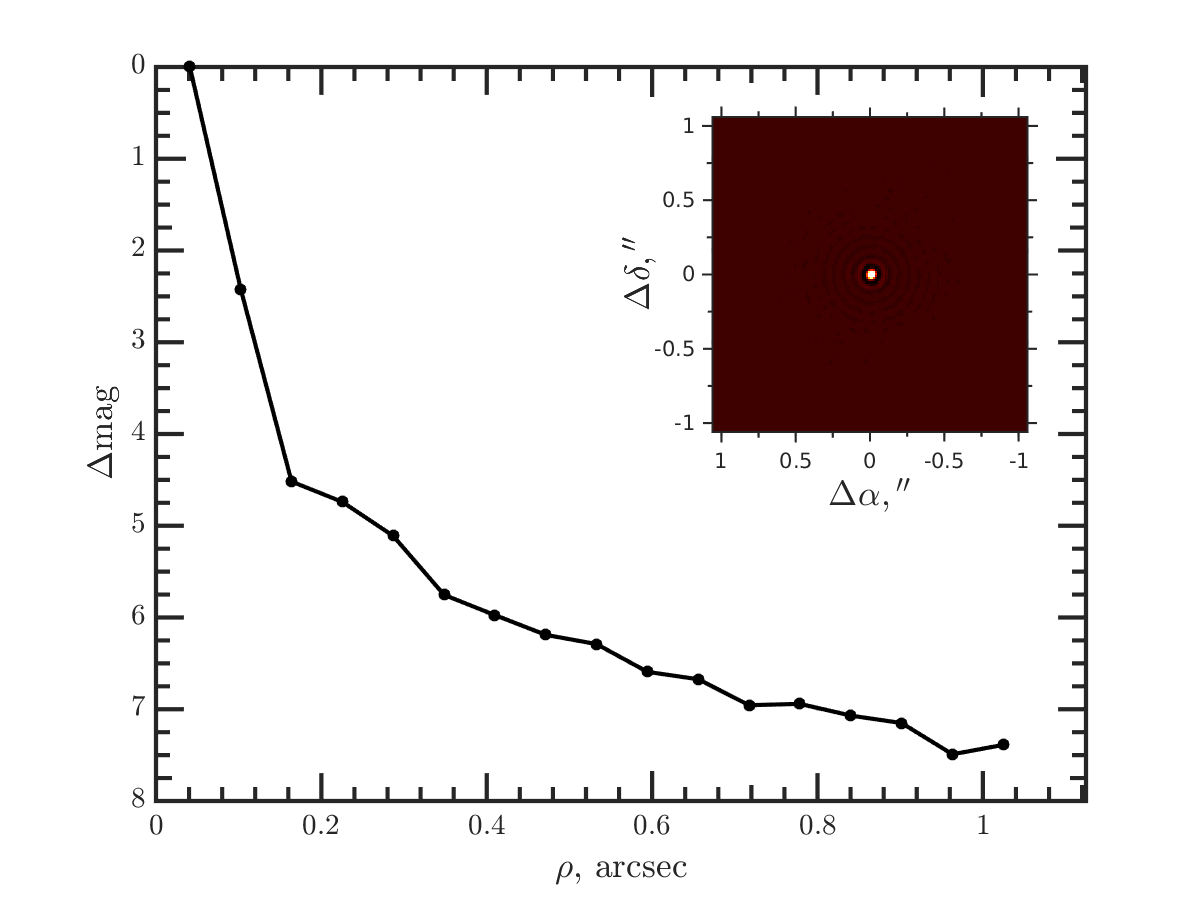}
\caption{Contrast curve for TOI-1135 from the speckle polarimeter instrument using a filter of 625\,nm. The inset figure shows the reconstructed 1$\times$1\arcsec{}$^2$. Any close companion was observed within 0.2\arcsec{} down to $\Delta$\,$=$\,4--5\,mag with respect to the target.
\label{fig:SAI}}
\end{figure}

We observed TOI-1135 on 28 November 2020 UT with the Speckle Polarimeter \citep{Safonov2017} on the 2.5\,m telescope at the Caucasian Observatory of Sternberg Astronomical Institute (SAI) of Lomonosov Moscow State University. Electron Multiplying CCD Andor iXon 897 was employed as a detector. The atmospheric dispersion compensator was active. Observations were conducted in a band centred at 625 nm and with a full width half maximum (FWHM) of 50 nm. The power spectrum was estimated from 4000 frames with 30 ms of exposure. The detector has a pixel size of 20.6 milliarcseconds (mas) on the sky, and the angular resolution was 63 mas. We did not detect any stellar companions brighter than $\Delta I_C$\,=\,4.8 mag and 7.2 mag at $\rho$\,=\,0.25\arcsec{} and 1.0\arcsec{}, respectively, where $\rho$ is the separation between the source and the potential companion (Fig.\,\ref{fig:SAI}). Thus, the companion detected in the NESSI speckle data is undetected in the SAI speckle data, suggesting that the companion has a redder colour, which we discuss later in Sect.\,\ref{sec:fap}.

\subsection{CRCAO transit photometry}
\label{sec:crcao}

The Carlson R. Chambliss Astronomical Observatory (CRCAO) is located on the campus of Kutztown University in Kutztown, Pennsylvania, USA. A full transit of TOI-1135\,b was observed from CRCAO in the $B$ filter for approximately six hours on 4 November 2020. A time series of 171 data images (120 s exposures) provided 55 minutes of pre-ingress and 67 minutes of post-egress baseline, and the short gap just after mid-transit resulted from a minor hardware malfunction. CRCAO employs a 0.61 m Ritchey-Chr\'etien optical telescope with a focal ratio of f/8. The imaging sensor (KAF-6303E) and telescope together produce a 19.5\arcmin$\times$13.0\arcmin field of view (FoV) with a plate scale of 0.76\arcsec pix$^{-1}$. Data reduction was performed with the \texttt{AstroImageJ} \citep[AIJ;][]{Collins2017} software package’s aperture photometry tool, using a 9.9\arcsec uncontaminated aperture. Due to TOI-1135's high declination, airmass was not required as a detrending parameter.

\subsection{Deep Sky West transit photometry}
\label{sec:dsw}

One full transit was observed on 10 December 2019 in the $g'$ band in the observatory Deep Sky West (DSW), near Rowe, New Mexico, USA. With an aperture of 0.5\,m, the CDK500 telescope is equipped with an Apogee Ultra 16m detector, which has an image scale of 1.09\arcsec pixel$^{-1}$, resulting in a 37\arcmin$\times$37\arcmin FoV. The exposure time of each image was 15\,s. The differential photometry was extracted with AIJ. The photometric aperture of 12\arcsec was adopted, which minimises the dispersion of the differential light curve.

\subsection{Peter van de Kamp Observatory transit photometry}
\label{sec:pvdk}

We observed a full transit of TOI-1135\,b on 25 November 2019 at the Peter van de Kamp Observatory (PvdKO) in the SDSS $i'$ filter. The PvdKO observatory is located in Swarthmore, Pennsylvania, USA. The telescope is a f/7.8 Ritchey-Chr\'etien with an aperture of 0.62\,m and an Apogee U16M 4K\,$\times$\,4K CCD camera. The pixel scale is 0.76\arcsec with 2$\times$2 binning, resulting in an FoV of 26\arcmin\,$\times$\,26\arcmin. The exposure time was 30\,s for each image. We employed an aperture of 8\arcsec to obtain the photometry of our star and several other stars in the FoV using the AIJ package.

\subsection{TTT transit photometry}
\label{sec:TTT}
Photometric data were obtained during three full transits on 22 December 2022 and on 7 and 15 January 2023 with the Two-Meter Twin Telescopes (TTT) facility as part of the key project observations during the commissioning of the telescope. The TTT facility is located at the Teide Observatory on the island of Tenerife, Canary Islands, Spain, and it currently has two telescopes, TTT1 and TTT2. They are both Ritchey-Chrétien optical systems with an aperture of 0.80\,m, an altazimuthal mount, and two f/6.85 Nasmyth foci each. During the observations, the TTT1 telescope was equipped with a 2K$\times$2K Andor iKon-L 936 camera with a back-illuminated 13.5 $\mu$m pixel$^{-1}$ BEX2-DD CCD sensor, resulting in an FoV of 17.3\arcmin$\times$17.3\arcmin and a plate scale of 0.51\arcsec pixel$^{-1}$. The TTT2 telescope was equipped with a QHY411M camera with a Sony IMX411 sCMOS sensor \citep{Alarcon2023} of 3.76 $\mu$m pixel$^{-1}$ and 151 megapixels. The effective FoV was 33.1\arcmin$\times$24.7\arcmin and the plate scale 0.14\arcsec pixel$^{-1}$. Data reduction was done using standard procedures, correcting for bias as well as dark and sky flat-fielding. The first of the three transits (22 December 2022) was observed simultaneously by both telescopes, with TTT1 in the $r'$ filter and TTT2 in the $g'$ filter and exposure times of 10 s and 15 s, respectively. The second transit (7 January 2023) was observed by TTT1 in the $r'$ and $g'$ filters, alternating filters on each image of the series, with an exposure time of 15 s in each band. The third transit (15 January 2023) was observed entirely in the $g'$ filter by TTT1 with an exposure time per image of 15 s. The differential photometry of each transit was extracted using the AIJ tool.

\subsection{SPECULOOS-North transit photometry}
\label{sec:spe}

We observed a full transit of TOI-1135\,b with the SPECULOOS-North/Artemis telescope on 22 December 2022 in the $z$ cut filter with an exposure time of 12\,s. The SPECULOOS-North/Artemis is a 1.0\,m Ritchey-Chretien telescope equipped with a thermoelectrically cooled 2K$\times$2K Andor iKon-L BEX2-DD CCD camera with a pixel scale of $0.35\arcsec$ and an FoV of 12\arcmin$\times$12\arcmin \citep{Burdanov2022}. The facility is located at the Teide Observatory in Tenerife, Canary Islands, Spain. It is a twin of the SPECULOOS-South located at ESO's Paranal Observatory in Chile \citep{Jehin2018Msngr,Delrez2018,Sebastian_2021AA} and SAINT-EX, located at the Sierra de San Pedro M\'artir in Baja California, M\'exico \citep{Demory_AA_SAINTEX_2020}. Data reduction and photometric measurements were performed using the {\tt PROSE}\footnote{\url{https://github.com/lgrcia/prose}} pipeline \citep{Garcia2021}.

\subsection{CARMENES spectroscopic observations}
\label{sec:carmrv}

We collected a total of 56 high-resolution spectra between 2 March and 16 May 2023 using the Calar Alto high-Resolution search for M dwarfs with Exoearths with Near-infrared and optical Echelle Spectrographs (CARMENES) instrument mounted on the 3.5\,m telescope located at the Calar Alto Observatory, Almer\'ia, Spain. These observations were conducted as part of the 23A-3.5-008 observing program (PI M.\,Mallorqu\'in). To accurately model the stellar activity, we implemented an observational strategy that aimed to obtain three to five spectra per stellar rotation period (5.1 days; Sect.\ \ref{sec:prot}). Finally, we were able to acquire an average of about four spectra per stellar rotation. The CARMENES spectrograph has two channels \citep{CARMENES, CARMENES18}, a visible (VIS) channel covering the spectral range of 520--960\,nm ($\mathcal{R}$\,=\,94600) and a near-infrared (NIR) channel covering the spectral range of 960--1710\,nm ($\mathcal{R}$\,=\,80400). One spectrum from each channel was ruled out because the instrumental drift correction was missing. Then, two and five spectra were removed due to their low signal-to-noise ratios (S/N\,$<$\,20) from the VIS and NIR channels, respectively. The final data sets contain 53 spectra in the VIS range and 49 spectra in the NIR range. The exposure time for these observations was 900\,s, resulting in an average S/N per pixel of 101 at 745\,nm and 83 at 1221\,nm. Detailed information on the performance of CARMENES, the data reduction, and wavelength calibration can be found in \cite{2016SPIE.9910E..0EC}, \cite{trifonov18}, and \cite{2018A&A...618A.115K}. Relative RVs and activity indicators, such as the chromatic index (CRX), differential line width (dLW), H$\alpha$ index, and the \ion{Ca}{ii} IR triple (IRT) were obtained using the software package {\tt serval}\footnote{\url{https://github.com/mzechmeister/serval}} \citep{2018A&A...609A..12Z}. The RV measurements were corrected for barycentric motion, secular acceleration, nightly zero-points, and for telluric lines as described in \citet{Nagel2023}. The typical dispersions of the RV measurements are $\sigma_{\mathrm{CARMENES\ VIS}}$\,$\sim$\,22.5 m\,s$^{-1}$ and $\sigma_{\mathrm{CARMENES\ NIR}}$\,$\sim$\,79.5 m\,s$^{-1}$, while the median uncertainties of the measured RVs are 8.9\,m\,s$^{-1}$ and 36.0\,m\,s$^{-1}$ for the VIS and NIR channels, respectively. Owing to the larger dispersion and error bars of the NIR RV data (a factor four and two with respect to the VIS RV data, respectively) and combined with the small amplitude signal expected for the planet, we decided to discard the NIR data in our subsequent analysis.

\begin{figure*}[ht!]
\includegraphics[width=1\linewidth]{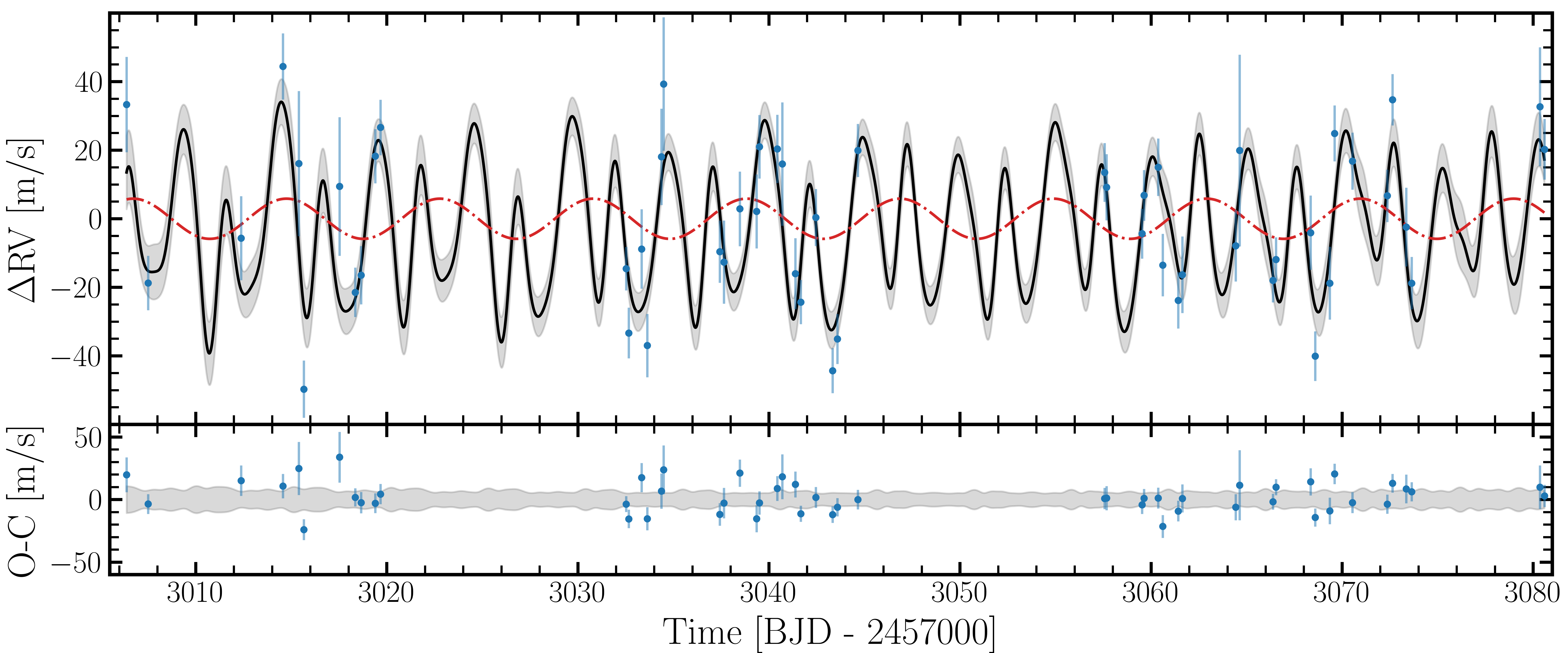}
\caption{CARMENES VIS RV data for TOI-1135 (blue dots). 
\textit{Top panel}: Combined model (black line) with its 1$\sigma$ level of confidence (grey shadow) and the Keplerian model for the planet (dashed red line). 
\textit{Bottom panel}: Residuals for the best fit.
\label{fig:RV_curve}}
\end{figure*}

The RV curve and its best-fit model (Sect.\ \ref{sec:joint} for details) are shown in Fig.\ \ref{fig:RV_curve}. Similar to the photometric TESS data, we examined the RV measurements to identify any epoch affected by flares by analysing the relative intensity of specific emission lines (H$\alpha$, \ion{Ca}{II}\,IRT, \ion{Na}{I}, and \ion{K}{I}) commonly associated with chromospheric activity. We compared these lines among all spectra to identify significant variations, but none were detected. Table \ref{tab:CARMV_RV} in the appendix provides the time stamps of the spectra in BJD$_{\mathrm{TDB}}$, the RVs measured using {\tt serval}, and the $1\sigma$ error bars for each RV measurement.

\section{Stellar properties}
\label{sec:properties}

\begin{table}
\caption{Stellar parameters of TOI-1135.}\label{tab:stellar_parameters}
\centering
\begin{tabular}{lcr}
\hline\hline
Parameter & Value & Reference\\
\hline
Name & HIP 62908 & Per97\\
     & TIC\,154872375 & TIC\\
\noalign{\smallskip}
$\alpha$ (J2016) & 12:53:35.1 & \textit{Gaia} DR3\\
$\delta$ (J2016) & +85:07:46.2 & \textit{Gaia} DR3\\
Sp. type & G0 & Nas46 \\
\noalign{\smallskip}
$\varpi$ [mas] & 8.758\,$\pm$\,0.011 & \textit{Gaia} DR3\\
$d$ [pc] & 113.920\,$\pm$\,0.145 & \textit{Gaia} DR3\\
$\mu_\alpha \cos{\delta}$ [mas yr$^{-1}$] & 16.973\,$\pm$\,0.015 & \textit{Gaia} DR3\\
$\mu_\delta$ [mas yr$^{-1}$] & 5.293\,$\pm$\,0.013 & \textit{Gaia} DR3\\
$\gamma$ [km\,s$^{-1}$] & --20.90\,$\pm$\,0.22 & \textit{Gaia} DR3\\
RUWE & $0.974$ & \textit{Gaia} DR3\\
\noalign{\smallskip}
$T_{\text{eff}}$ [K] & 6122\,$\pm$\,15 & This work \\
$\log{g}$ [cgs] & 4.56\,$\pm$\,0.02 & This work \\
{[Fe/H]} [dex] & 0.00\,$\pm$\,0.02 & This work \\
\noalign{\smallskip}
$M_{\star}$ [$M_{\odot}$] & 1.125\,$\pm$\,0.032 & This work \\
$R_{\star}$ [$R_{\odot}$] & 1.160\,$\pm$\,0.029 & This work \\
$L_{\star}$ [$L_{\odot}$] & 1.702\,$\pm$\,0.068 & This work\\ 
$v\sin i$ [km\,s$^{-1}$] & 10.4\,$\pm$\,0.1 & This work\\
$v_{\text{macro}}$ [km\,s$^{-1}$] & 5.5\,$\pm$\,0.1 & This work\\
$P_{\text{rot}}$ [d] & 5.13\,$\pm$\,0.27 & This work\\
$U$ [km s$^{-1}$] & 16.69\,$\pm$\,0.03 & This work\\
$V$ [km s$^{-1}$] & --8.95\,$\pm$\,0.04 & This work\\
$W$ [km s$^{-1}$] & --13.84\,$\pm$\,0.03 & This work\\
Gal. population & Young disc & This work\\
Age [Myr] & 125--1000 & This work\\
$NUV$ [mag] & 14.219\,$\pm$\,0.006 & GALEX\\
$B_{p}$ [mag] & 9.695\,$\pm$\,0.003 & \textit{Gaia} DR3\\
$V$ [mag] & 9.57\,$\pm$\,0.02 & Tycho-2\\
$G$ [mag] & 9.419\,$\pm$\,0.003 & \textit{Gaia} DR3\\
$R_{p}$ [mag] & 8.974\,$\pm$\,0.004 & \textit{Gaia} DR3\\
$J$ [mag] & 8.458\,$\pm$\,0.026 & 2MASS\\
$K$ [mag] & 8.192\,$\pm$\,0.019 & 2MASS\\
EW(Li) [m\AA] & 82$\pm$10 & This work\\
\noalign{\smallskip}
\hline
\end{tabular}
\tablebib{
Per97: \citet{Perryman1997}; 
TIC: \citet{Stassun2019};
Nas46: \citet{Nassau1946};
\textit{Gaia} DR3: \cite{gaia2016, gaiadr3}; 
Tycho-2: \cite{hog_2000}; 
\textit{GALEX}: \citet{bianchi2017};
2MASS: \citet{skrutskie2006}.
}
\end{table}

Using the CARMENES stellar template computed by \texttt{serval}, after combining all 53 individual spectra, we obtained the stellar atmospheric parameters of TOI-1135\,by means of the {\sc SteParSyn} code\footnote{\url{https://github.com/hmtabernero/SteParSyn/}} \citep{tab22}. This code implements the spectral synthesis method with the {\tt emcee}\footnote{\url{https://github.com/dfm/emcee}} Python package \citep{emcee} in order to retrieve the stellar atmospheric parameters. We employed a grid of synthetic spectra computed with the Turbospectrum \citep{Plez2012} code, the MARCS stellar atmospheric models \citep{Gustafsson2008}, and the atomic and molecular data of the $Gaia$-ESO line list \citep{Heiter2021}. We employed a set of \ion{Fe}{i,ii} features well suited to analysing the FGKM stars listed in \citet{tab22}. Thus, we retrieved the following parameters: $T_{\rm eff}$\,$=$\,6122\,$\pm$\,15\,K; $\log{g}$\,$=$\,4.56\,$\pm$\,0.02 dex; [Fe/H]\,$=$\,0.00\,$\pm$\,0.02 dex; and $\varv \sin{i}$\,$=$\,10.4\,$\pm$\,0.1 km~s$^{-1}$ and $\varv_{\rm macro}$\,$=$\,5.5\,$\pm$\,0.1~km~s$^{-1}$.\\

We estimated the luminosity of TOI-1135\,by integrating the observed fluxes from the UV-optical to mid-infrared using VOSA \citep{bayo08}, including the Galaxy Evolution Explorer (GALEX; \citealp{bianchi2017}), the Hipparcos \citep{Perryman1997}, the Panoramic Survey Telescope and Rapid Response System (Pan-STARRS; \citealp{Chambers2016}), the Hubble Source Catalog \citep{Whitmore2016} catalogues, {\sl Gaia} DR3 \citep{gaia2016, gaiadr3}, the Sloan Digital Sky Survey (SDSS; \citealp{york_2000}), the Two Micron All-Sky Survey (2MASS; \citealp{skrutskie2006}), the Javalambre Physics of the Accelerating Universe Astrophysical Survey (J-PAS; \citealp{Cenarro2019}), the Javalambre Photometric Local Universe Survey (J-PLUS; \citealp{Dupke2019}), the Johnson $UBVR$ photometry \citep{Ducati2002}, and the Wide-field Infrared Survey Explorer (WISE; \citealp{wright_2010}). We used BT-Settl (CIFIST) models \citep{baraffe15} to reproduce the spectral energy distribution (SED) of the star and to extrapolate to bluer and longer wavelengths. We obtained for TOI-1135 a luminosity of 1.702\,$\pm$\,0.068 $L_{\odot}$. From the estimated effective temperature and luminosity, and using the Stefan-Boltzman relation, we derived a radius of 1.160\,$\pm$\,0.029 $R_{\odot}$. Assuming that the star is on the main sequence, which is expected for its age (see Sect.\,\ref{sec:age}), and using the empirical mass-luminosity relations for solar-type stars from \citet{eker18}, we determined a mass of 1.125\,$\pm$\,0.032 $M_{\odot}$, which includes the error due to the dispersion of mass-luminosity fit. The summary of the main stellar parameters of TOI-1135 can be found in Table\ \ref{tab:stellar_parameters}.\\

No wide companions of TOI-1135 have been reported in the literature. We searched for common proper motion companions to TOI-1135 using the $Gaia$ DR3 catalogue and examined objects within a radius of 1 degree, which corresponds to a physical separation close to 2 pc at the distance of the star. To narrow down the search, we applied a restriction on the parallax values, specifically within the range of 8.25--9.25 mas, which encompasses the parallax of TOI-1135 as listed in Table \ref{tab:stellar_parameters}. The query returned 10 stars. However, the proper motions of these stars differ greatly from those of TOI-1135, indicating that they are most likely unrelated. Therefore, no wide common proper motion companions were identified within the $Gaia$ catalogue within this restricted search range. 

\subsection{Rotation period and spectral stellar activity indicators}
\label{sec:prot}

Active regions such as spots and faculae on the surface of F, G, K, and M stars exhibit a periodic behaviour as the star rotates. Furthermore, these active regions can shift across the stellar surface and appear and disappear as the star rotates, resulting in quasi-periodic (QP) photometric and spectroscopic variability. By performing a frequency analysis of their light curves, RV, and spectral stellar activity indicators, it is possible to determine the rotation period of these stars and their activity levels. 

\begin{figure}
\includegraphics[width=9.cm]{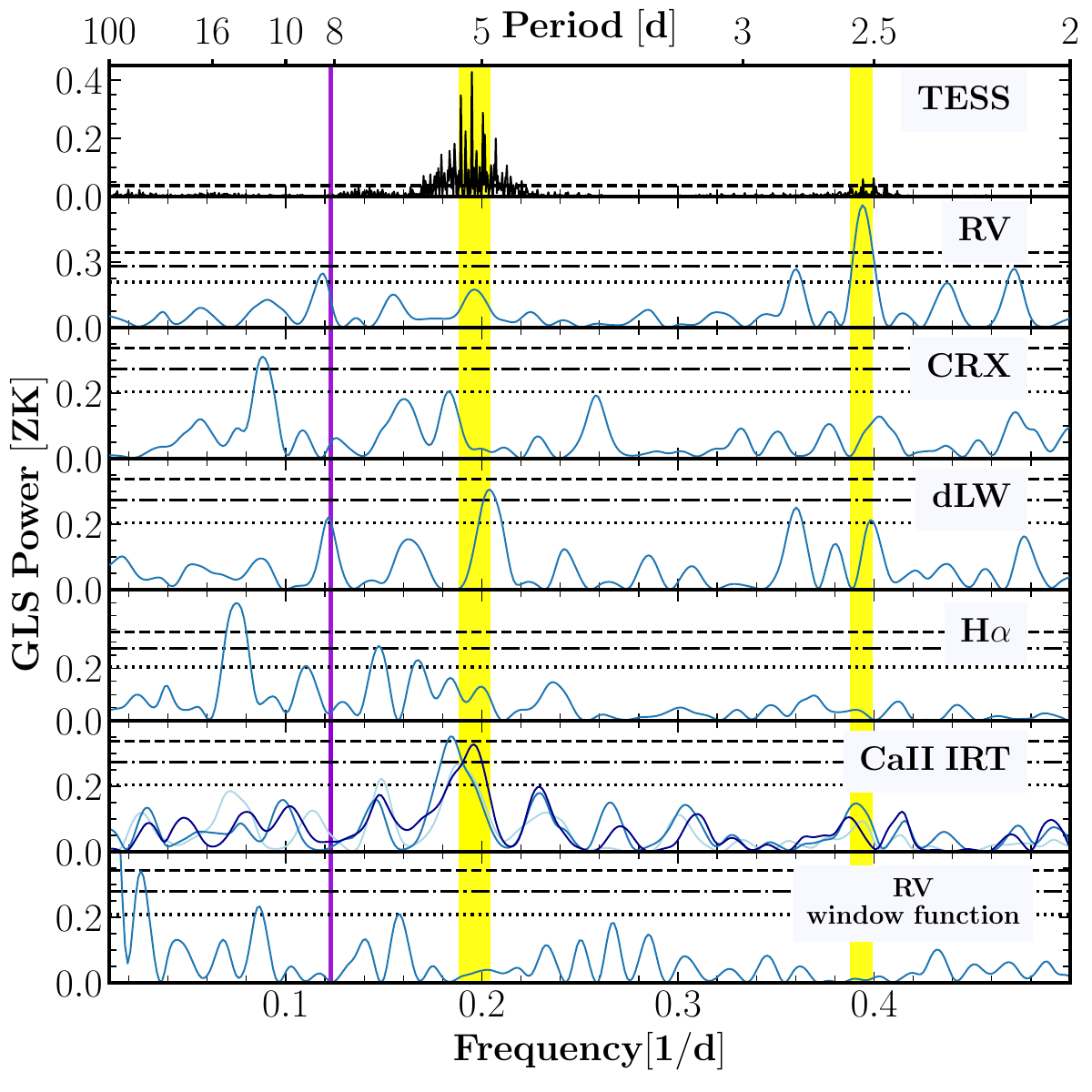}
\caption{GLS periodograms for TESS light (black line) and CARMENES VIS (blue lines) data set analysis of the photometric and spectroscopic data. In the sixth panel (from top to bottom), the \ion{Ca}{ii} IRT is plotted in light (1), medium (2), and dark (3) colours. The stellar rotation period and its first harmonic are shown as two vertical yellow bands centred at 0.196~d$^{-1}$ (5.1 d) and 0.392~d$^{-1}$ (2.6 d). The vertical purple line indicates the orbital period of the transiting planet. The horizontal dashed blue lines correspond to the FAP levels of 0.1\%, 1\%, and 10\%. However, the TESS panel only includes the 0.1\% FAP multiplied by a factor of 10 for clarity.
\label{fig:GLS_Prot}}
\end{figure}

Figure\ \ref{fig:GLS_Prot} shows the generalised Lomb-Scargle (GLS; \citealp{Zechmeister2009}) periodograms for the combination of TESS's sectors as well as for the RV, the spectral stellar activity indicators, and the window function of the CARMENES data. The panels that include TESS and \ion{Ca}{ii} IRT data show that the most significant signal is slightly more than five days (it is also observed in dLW but with less significance), while in the RV panel, the most significant signal is shown at half that period. From these periodograms combined with the TESS light curves (Fig.\ \ref{fig:LC_TESS}), where a very high cadence is available covering more than 200 days, we conclude that the stellar rotation period is 5.13\,$\pm$\,0.27 days.

In addition to the signals associated with the rotation period, we identified a signal in the H$\alpha$ panel with a FAP of less than 0.1\% at approximately 13 days, although its origin remains unclear. Furthermore, we performed an analysis of correlation using Pearson's $r$ coefficient to investigate potential correlations between the CARMENES RV data and the activity indicators, but we did not detect any significant correlations.

\subsection{Age}
\label{sec:age}

The relatively short stellar rotation period ($P_{\text{rot}}$\,$=$\,5.13\,$\pm$\,0.27\,d) of TOI-1135 may be indicative of youth. Young stars often exhibit fast rotation because they retain angular momentum from their formation process. This high-rotation star shows high levels of chromospheric activity and a larger presence of spots on its surface. As these stars age, they undergo a process of rotational braking through magnetic interactions \citep{Gallet13, Gallet15}, causing their rotation to gradually slow down. By studying various age indicators, such as the gyrochronology, the NUV excess, the kinematics, and lithium equivalent width, we can restrict the age of the TOI-1135 star.

\subsubsection{Gyrochronology}

\begin{figure}[ht!]
\includegraphics[width=1\linewidth]{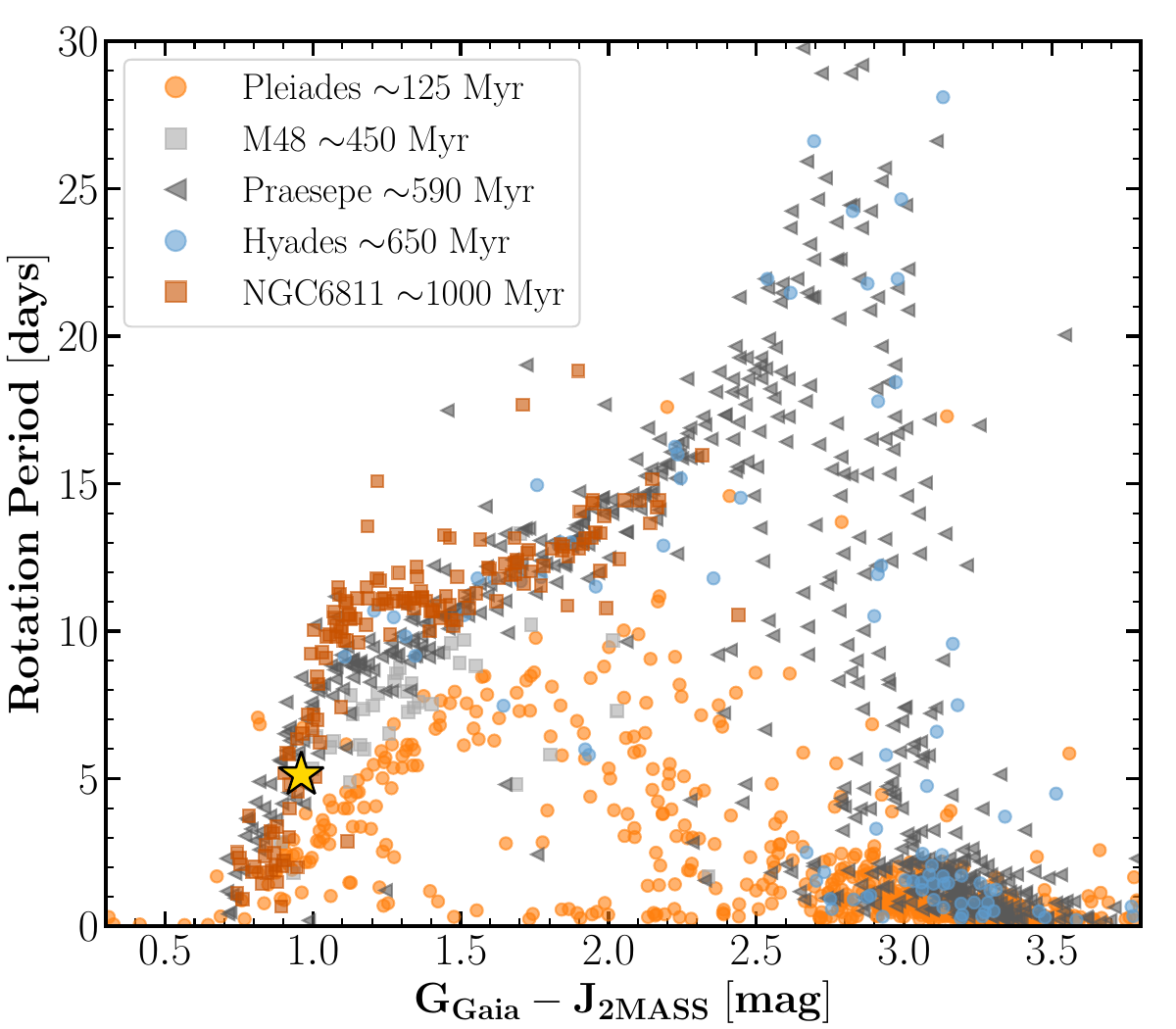}
\caption{Distribution of rotation stellar periods against the $G-J$ colour for Pleiades ($\sim$125 Myr; \citealp{Rebull2016}), M48 ($\sim$450 Myr; \citealp{Barnes2015}), Praesepe ($\sim$590 Myr; \citealp{Douglas2017}), Hyades ($\sim$650 Myr; \citealp{Douglas2019}), and NGC6811 ($\sim$1000 Myr; \citealp{Curtis2019}) clusters. TOI-1135 is represented as a gold star. 
\label{fig:prot_age}}
\end{figure}

Figure\ \ref{fig:prot_age} depicts the distribution of stellar rotation periods as a function of colour $G-J$ for several stellar clusters at different ages, namely, the Pleiades ($\sim$125 Myr; \citealp{Rebull2016}), M48 ($\sim$450 Myr; \citealp{Barnes2015}), Praesepe ($\sim$590 Myr; \citealp{Douglas2017}), Hyades ($\sim$650 Myr; \citealp{Douglas2019}), and NGC6811 ($\sim$1000 Myr; \citealp{Curtis2019}). In this diagram, TOI-1135 lies above the Pleiades sequence but on top of the M48, Praesepe, Hyades, and NGC6811 sequences. Therefore, gyrochronology indicates that TOI-1135 has an age older than $\sim$125 Myr and equal to or younger than $\sim$1000 Myr.

\subsubsection{NUV excess} 

\begin{figure}[ht!]
\includegraphics[width=1\linewidth]{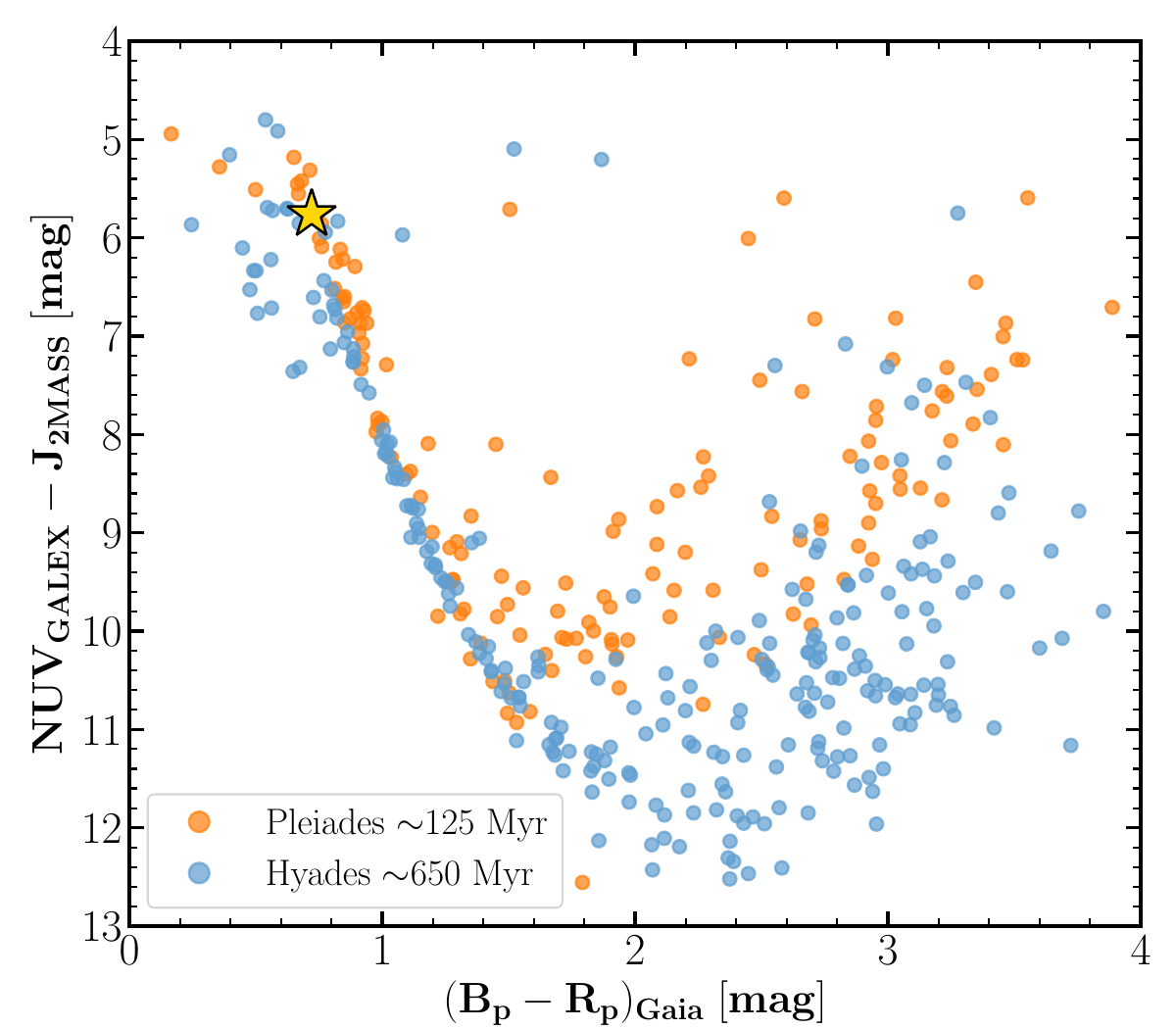}
\caption{Colour distribution of $NUV-J$ as a function of $B{p}-R{p}$ colour for the Pleiades ($\sim$125 Myr; \citealp{olivares2018}) and Hyades ($\sim$650 Myr; \citealp{roser2019}) clusters. TOI-1135 is represented as a gold star.
\label{fig:nuv_age}}
\end{figure}

We also used the NUV emission as an indicator of youth, as its strength decreases with stellar age \citep{findeisen2011}. \cite{shkolnik2011} and \cite{rodriguez2011} used either the flux ratio $F_{NUV}/F_{J}$ or the $m_{NUV}$\,--\,$m_{J}$ colour to identify young stars. The $NUV$ magnitude comes from the GALEX all-sky catalogue \citep{bianchi2017}. However, this method does not provide a clear age determination for a solar-type star like TOI-1135\, because these criteria are better suited for low-mass stars. In the $NUV$--$J$ versus $B_p$\,--\,$R_p$ colour plot (Fig.\ \ref{fig:nuv_age}), TOI-1135 is located on top of the sequences of the Pleiades ($\sim$125 Myr) and Hyades ($\sim$650 Myr) clusters, which is the same as the stars in the main sequence for solar-type stars. However, it is not clear that TOI-1135 has a UV excess, and therefore, a clear age constraint cannot be obtained with this method.

\subsubsection{$UVW$}

\begin{figure}[ht!]
\includegraphics[width=1\linewidth]{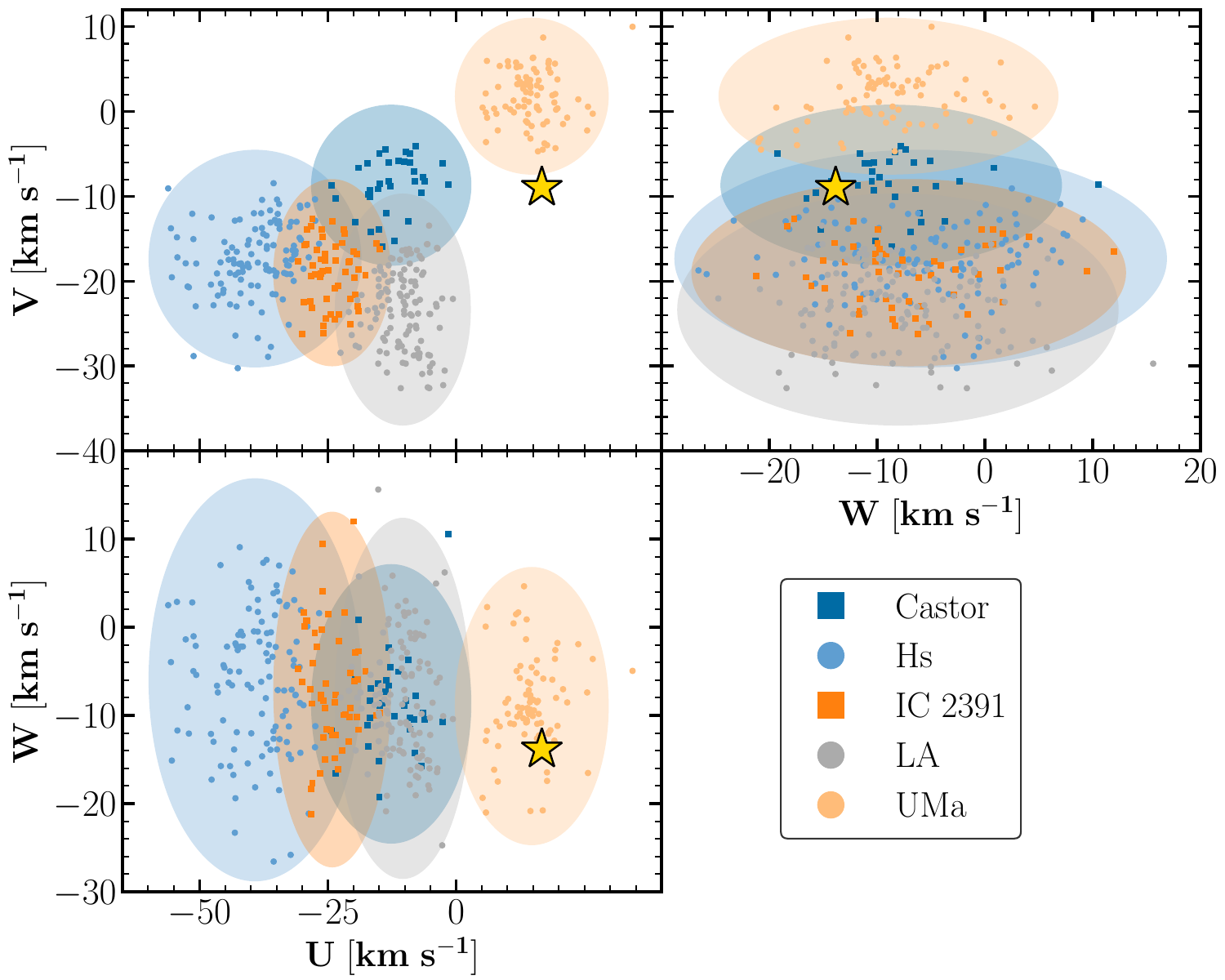}
\caption{Diagram of $UVW$ velocities for TOI-1135 (gold star). The members of the Castor moving group (200--400\,Myr), the Hyades supercluster (Hs; 600--800 Myr), the IC 2391 supercluster (35--55 Myr), the Local Association (LA; 10--300 Myr), and the Ursa Major group (UMa; $\sim$400 Myr) from \cite{montes2001b} are included. The ellipses represent the 3$\sigma$ values of the $UVW$ for each group.
\label{fig:UVW}}
\end{figure}

If TOI-1135\,belongs to a known moving group with a well-determined age, we can assign the age of that group to this star. We studied the kinematics and $UVW$ galactocentric space velocities of well-known clusters, moving groups, and star-formation regions to figure out whether TOI-1135 is associated with such a region in an attempt to narrow down its age. Space velocities provide information about the motion of objects in the Galaxy. Using the astrometric data from the \textit{Gaia} mission \citep{gaia2016, gaiadr3}, we calculated the $UVW$ velocities following the method described in \cite{johnson1987}. In this convention, $U$ is positive towards the Galactic centre, $V$ is positive in the direction of Galactic rotation, and $W$ is positive towards the north galactic pole. The $UVW$ velocities of TOI-1135 are listed in the Table\ \ref{tab:stellar_parameters} and plotted in Fig.\ \ref{fig:UVW} along with a compilation of young moving group members from \citet{montes2001b}. The velocities of TOI-1135 have young kinematics but are not consistent with any of the young moving groups depicted in Fig.\ \ref{fig:UVW} at 3$\sigma$.

\subsubsection{Lithium equivalent width}

\begin{figure}[ht!]
\includegraphics[width=1\linewidth]{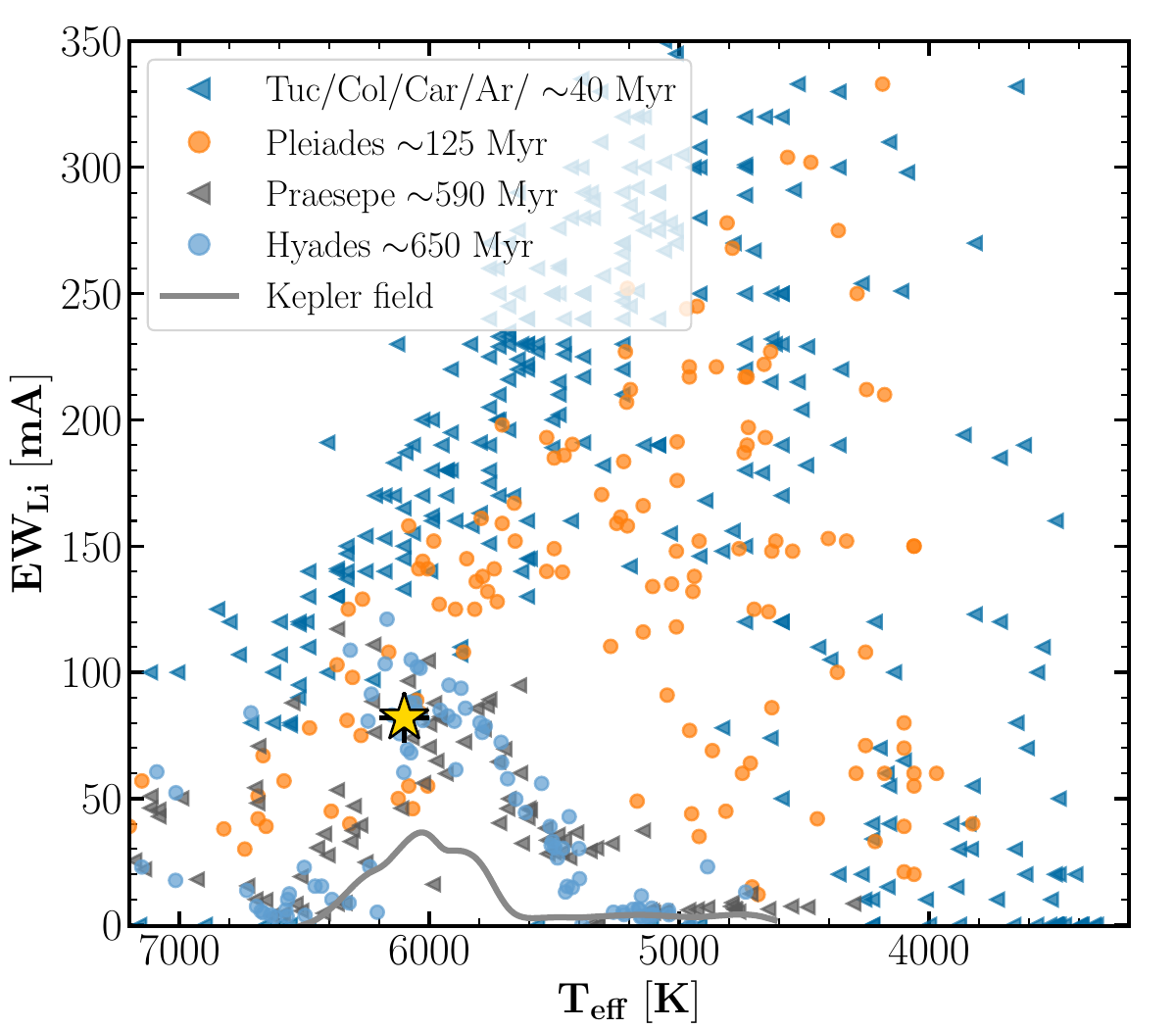}
\caption{Distribution of the Li{\small{I}} pseudo-equivalent widths as a function of effective temperature ($T_{\text{eff}}$). The figure includes members of similar-aged moving groups \citep[$\sim$40 Myr; TucHor, Columba, Carina, Argus;][]{daSilva2019}, Pleiades \citep[125 Myr;][]{Bouvier2018}, the Praesepe \citep[635$\pm$25 Myr;][]{Cummings2017}, and Hyades \citep[670$\pm$25 Myr;][]{Cummings2017}, which are shown as dark blue, orange, grey, and blue symbols, respectively. Moreover, the Kepler field stars from \citep{Bouvier2018} are shown with a grey solid line. The \ion{Li}{i} EW of TOI-1135 is represented as a gold star.
\label{fig:EW_Li}}
\end{figure}

The depth of the lithium resonance feature ($^{7}$Li) is commonly used as an age estimator in solar-type stars. The lithium atoms are rapidly burned into heavier elements within the first few hundred million years, causing the strength of the absorption feature to decrease with stellar age. In the case of TOI-1135, we have measured the equivalent width (EW) of the Li feature in the co-added CARMENES template spectrum of TOI-1135\,by fitting a Gaussian profile that includes the Li doublet (6707.76\,\AA\ and 6707.91\,\AA). We measured an EW of 0.082\,$\pm$\,0.010\,\AA\ (Fig.\ \ref{fig:EW_Li}), whose value is consistent with members of the Praesepe and Hyades (610--695 Myr; \citealp{Cummings2017}) clusters and clearly larger than those of Kepler field stars from \citet{Bouvier2018}. However, the EW is smaller than groups younger than a few tens of million years, indicating that the star has not fully preserved its lithium and is in the process of destroying it.

Taking into consideration the age ranges derived from the previous age indicators, we adopted a conservative age range of 125--1000 Myr. Therefore, we considered this star as young, in terms of planet formation and evolution mechanisms.

\section{Analysis}
\label{sec:analysis}

\subsection{Transit search}
\label{sec:trsearch}

Searching for transits in light curves of active and/or young stars is especially difficult because it requires precise modelling of the stellar activity. This presents QP variations whose timescales are on the order of days, the same as those of the orbital periods of the planets (Fig.\ \ref{fig:LC_TESS}). Therefore, the modelling of the activity must be done carefully, adequately capturing this type of activity without removing planetary transits. The Gaussian process regression \citep[GP;][]{Rasmussen2006} provides sufficiently flexible functions to model the variations in amplitude and the QP behaviour of light curves. Specifically, we used the double simple harmonic oscillator (dSHO) kernel implemented in the \texttt{celerite} package \citep{celerite}, which has been widely used in the literature to model the photometric stellar activity \citep{david18a, mann20, newton21, toff21, Suarez-Mascareno2022, Mallorquin2023}, defined as 

\begin{equation}
\begin{aligned}
k_{\mathrm{dSHO}}(\tau) = & \  k_{\mathrm{SHO}}(\tau; \eta_{\sigma_{1}}, \eta_{L_{1}}, \eta_{P}) + k_{\mathrm{SHO}}(\tau; \eta_{\sigma_{2}}, \eta_{L_{2}}, \eta_{P}/2) \\
               = & \ \eta_{\sigma_{1}}^{2}  e^{-\frac{\tau}{\eta_{L_{1}}}}  \left[ \cos \left(\eta_{1} \frac{2\pi\tau}{\eta_{P}} \right) + \eta_{1} \frac{\eta_{P}}{2\pi \eta_{L_{1}}} \sin \left(\eta_{1} \frac{2\pi\tau}{\eta_{P}} \right) \right] \\
             & + \eta_{\sigma_{2}}^{2}  e^{-\frac{\tau}{\eta_{L_{2}}}}  \left[ \cos \left(\eta_{2} \frac{4\pi\tau}{\eta_{P}} \right) + \eta_{2} \frac{\eta_{P}}{4\pi \eta_{L_{2}}} \sin \left(\eta_{2} \frac{4\pi\tau}{\eta_{P}} \right) \right],
\end{aligned}
\label{eq:dsho}
\end{equation}

where $\tau$\,$\equiv$\,$|t_i - t_j|$ represents the time difference between two data points; $\eta$ is defined as $|1 - (2\pi \eta_{L}/\eta_{P})^{-2} |)^{1/2}$; and $\eta_{\sigma_{i}}$, $\eta_{L_{i}}$, and $\eta_{P}$ are the hyperparameters that quantify the amplitude of the covariance, the decay timescale, and the period of the fundamental signal. Therefore, we related the hyperparameters $\eta_{P}$ as the stellar rotation of the star, $\eta_{L_{i}}$ as the evolution timescale of the active regions, and $\eta_{\sigma_{i}}$ as the variations of the amplitudes in the light curve. It is important to note that this kernel definition is valid only if $\eta_{P}$\,$<$\,$2\pi \eta_{L}$. This assumption is reasonable for young stars since they typically exhibit a dominant periodic behaviour in their activity. Our goal is to create smooth functions that model stellar activity but not possible transits. For this, we adapted the hyperparameters to the dominant induced stellar activity scales seen in the light curve: the stellar activity due to rotation. Furthermore, we also included an instrumental offset ($\gamma_{\mathrm{TESS}}$) as well as a jitter ($\sigma_{\mathrm{jit,TESS}}$) term added in quadrature to the error bars. We set normal priors to the $\eta_{\sigma_{i}}$ and the $\eta_{P}$ hyperparameters to $\sim$3.5 ppt and 5.1 days, which corresponds to the dispersion observed in the TESS data and the rotation period of the star, respectively. However, we used uniform priors for the $\eta_{L_{i}}$ hyperparameters with a minimum of 28 days (the baseline of one sector of TESS with the aim of modelling only the long-term scales). Lastly, we fixed the $\sigma_{\mathrm{jit,TESS}}$ parameter to $\sim$6 ppt, which is the observed transit depth of TOI-1135\,b, in order to avoid modelling such transit depths. The parameter space was explored with two different sampling algorithms for the Markov chain Monte Carlo (MCMC) process. First, we employed an affine-invariant ensemble sampler \citep{goodman10} implemented in the \texttt{emcee} code \citep{emcee}. This sampler allows for efficient exploration of the parameter space by generating a diverse ensemble of walkers. Additionally, we employed the \texttt{dynesty} algorithm \citep{dynesty},\footnote{\url{https://github.com/joshspeagle/dynesty}} which is based on nested sampling \citep{skilling04}. The \texttt{dynesty} algorithm provides an alternative approach to explore the parameter space by iteratively updating a set of live points used to estimate the evidence and posterior distributions. We found that both sampling algorithms show similar posteriors, so from hereon we only show the results obtained with \texttt{emcee}.

We searched for transits using the box least square periodogram (BLS; \citealp{kovacs2002, hartman2016}) and found a signal with a period of 8.027 days as the most significant signal in the light curve that agrees with the period alerted by the TSO. To search for additional transits, we then masked out this signal and applied the BLS algorithm iteratively. However, no significant additional transit signals were found beyond the first one. We further examined the light curves visually, searching for isolated transits, and we did not identify any variations of this type.

\subsection{Transit analysis}
\label{sec:trchr}

We used the TESS and the ground-based multi-band photometric transit follow-up to study the transit chromaticity. To do this, we created a transit model assuming circular orbits with the \texttt{PyTransit}\footnote{\url{https://github.com/hpparvi/PyTransit}} \citep{Parviainen2015} package, where we used the following planetary parameters: the time-of-transit centre ($T_c$), the orbital period of the planet ($P$), the planet-star radius ratio ($R_p/R_{\star}$), the orbital semi-major axis divided by the stellar radius ($a/R_\star$), the impact parameter ($b$), the eccentricity ($e$), and the argument of periastron ($\omega$). The parameter $a/R_\star$ was re-parameterised using Kepler's third law, which under the assumption that $M_p/M_\star$\,$\ll$\,1, only depends on the orbital period of the planet and the radius and mass of the star ($R_\star$, $M_\star$). The radius and mass of the star were introduced in the fit as fixed parameters (according to the central values of Table\,\ref{tab:stellar_parameters}), while the derived parameters that depend on the radius and mass of the star (such as $R_p$, $a/R_\star$ or $a$) were calculated a posteriori using the corresponding error propagation. In addition, we took into account the variations in the shape of the transit due to limb darkening, adopting the parameterisation proposed by \cite{kipping02013} for the quadratic limb darkening coefficients (from $u_1$, $u_2$ to $q_1$, $q_2$). These coefficients are different for each filter, and the initial values were calculated using the \texttt{PyLDTk} \citep{Parviainen2015b} tool. As in the previous section, we used the \texttt{emcee} code to sample the parameter space. For $T_c$ and $P$, we used normal priors around the BLS solution, while for the $R_p/R_{\star}$ and $b$ parameters, we used uniform priors. In addition, for each instrument, we added an instrumental offset and a jitter term as uniform priors. In the case of transit photometry, the jitter term is enough to capture the extra noise, but it is not so in the case of TESS photometry, where it is necessary to model stellar activity. Thus, we used the same GP kernel, the dSHO, to model the activity in TESS data as in Sect.\ \ref{sec:trsearch}, where the jitter term is a free parameter. In summary, we fit all the photometric data sets with transits sharing all parameters except $\gamma$ and $\sigma_{\mathrm{jit}}$, which are different for each instrument, and $R_p/R_{\star}$ (to study the transit chromaticity), $q_1$, and $q_2$, which are independent for each filter.

\begin{figure}[ht!]
\includegraphics[width=1\linewidth]{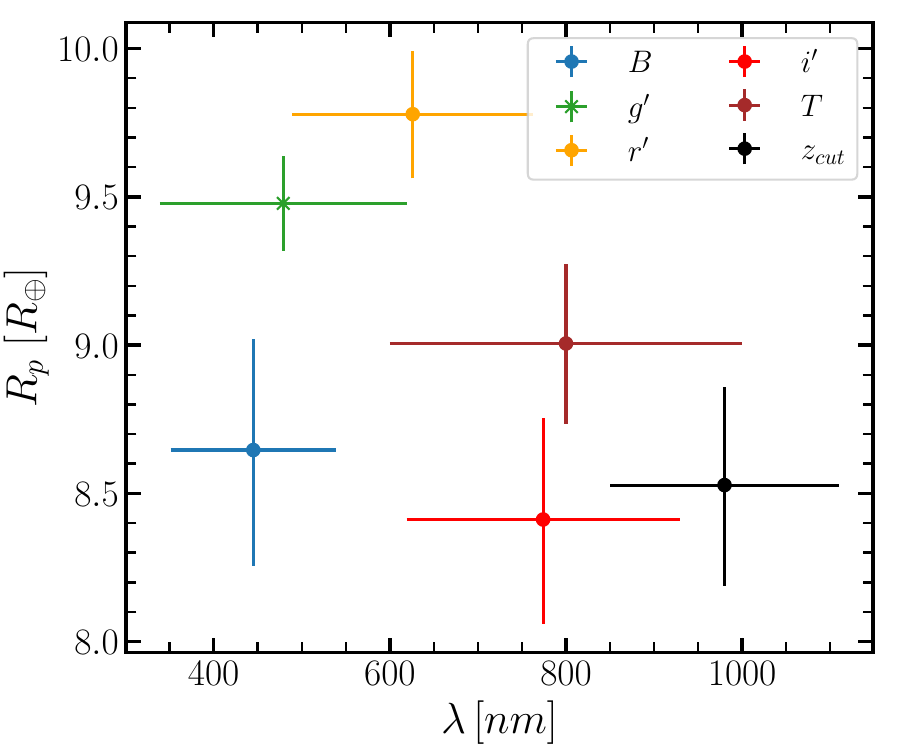}
\includegraphics[width=1\linewidth]{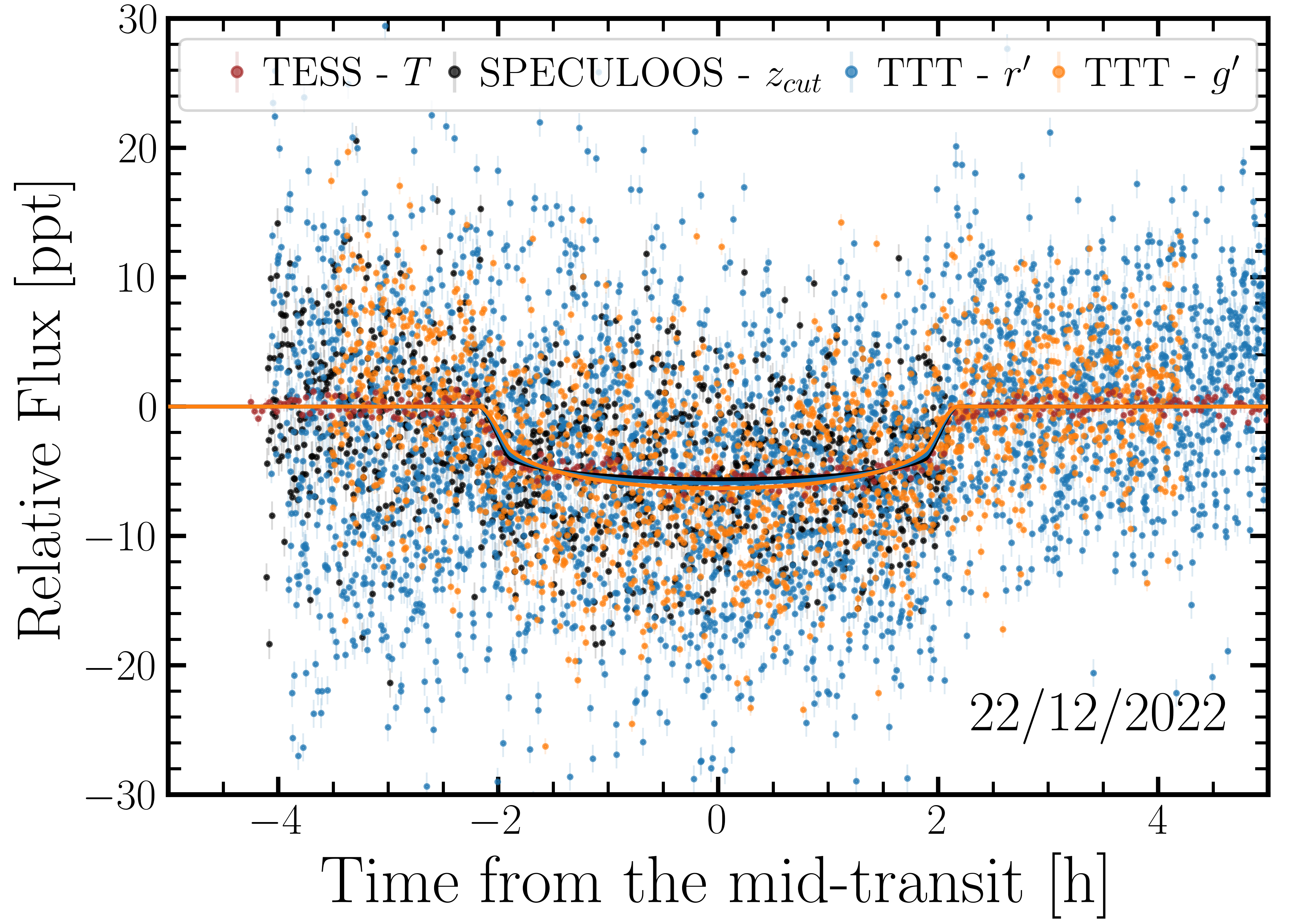}
\caption{Transit chromaticity of TOI-1135\,b. \textit{Top panel}: Planetary radius as a function of wavelength. The figure shows that there is no clear chromaticity in the radius of TOI-1135\,b in the range of 400--1000 nm. \textit{Bottom panel}: Phased-folded transit taken on 22 December 2022 in $T$, $g'$, $r'$, and $z_{cut}$ filters. The coloured lines show the best transit model in each filter.
\label{fig:trchr}}
\end{figure}

In Fig.\ \ref{fig:trchr}, the top panel shows the wavelength dependence of the measured radius. Within the precisions obtained, there does not seem to be a clear chromaticity in the transit of TOI-1135\,b. Additionally, the bottom panel of the figure shows the transit that occurred on 22 December 2022, which was simultaneously observed in the $g'$ (TTT\,2), $r'$(TTT\,1), $z_{cut}$ (SPECULOOS), and $T$ (TESS sector 59) filters. Therefore, in subsequent transit analyses, we use a single parameter of $R_p/R_{\star}$.

\subsection{Radial velocity analysis}
\label{sec:rvanalysis}

Before performing any RV analysis, we investigated if some of our data were acquired during some of the transits. The Rossiter-McLaughlin (RM) effect \citep{rossiter, mclaughlin} can be significantly high on giant planets. We estimated the expected amplitude in RV for TOI-1135\,b following \citet{gaudi07}. We calculated a semi-amplitude of 50.2\,$\pm$\,10.6\,m\,s$^{-1}$, a value comparable to the rms of the CARMENES data sets. The estimated RM semi-amplitude is well in agreement to that detected during the in-transit observations (Orell-Miquel et al., in prep.). However, all the CARMENES data were taken out-of-transit. 

In Sect.\ \ref{sec:prot} we computed GLS periodograms over the RV and activity indicators of CARMENES VIS data (Fig.\ \ref{fig:GLS_Prot}). The only significant signal in the RV panel is related to the first harmonic of the stellar rotation period ($\sim$2.6\,d). Therefore, the signal of the transiting planet (purple line) is not present in either the GLS RV periodogram or in the activity indicators. In active and young stars such as TOI-1135, this is an expected behaviour since stellar activity is usually significantly larger than the expected amplitude for the planet, so the signal of the planet may be hidden until the stellar activity is modelled.

We performed an RV-only fit following the same procedure as in \citet{Mallorquin2023, Mallorquin2023b}, to determine the best model to fit the stellar activity using three different approaches. The first model uses a jitter term ($\sigma_{\mathrm{jit,RV}}$), added in quadrature to the error bars of the RV measurements, to take into account possible additional noise not captured by the model selected. The second model, in addition to the jitter term, incorporates two sinusoidal functions centred on the stellar rotation period and on half of the rotation period. By including these sinusoidal functions, our aim was to capture the periodic variations introduced by the stellar rotation observed in Fig.\ \ref{fig:GLS_Prot}. Lastly, the third model uses a jitter term and a QP kernel introduced by \citet{aigrain12}: 

\begin{equation}
k_{\mathrm{QP}}(\tau)  =  \eta_{\sigma}^2 \exp \left[ -\frac{\tau^2}{2\eta_{L}^2} -\frac{\sin^2{ \left( \frac{\pi \tau}{\eta_{P}} \right)}}{2\eta_{\omega}^2} \right], \\
\label{eq:qp}
\end{equation}

where $\tau$, $\eta_{\sigma}$, $\eta_{L}$, and $\eta_{P}$ are defined as in Eq.\ \ref{eq:dsho}, and $\eta_{\omega}$ acts as a balance between the periodic and non-periodic component of the kernel. This kernel allows for a more flexible modelling of the stellar activity by incorporating QP variations in addition to the random noise captured by the jitter term. These models allowed us to explore different representations of the stellar activity and to assess their effectiveness in capturing the observed RV variations. In the RV model, in addition to the activity model, a Keplerian associated with the transiting planet is also included, which was tested against the same activity models without including the Keplerian. This allowed us to study how significant it is to include the planet or not in our RV data. We modelled the Keplerian signal of the transiting planet with \texttt{RadVel}\footnote{\url{https://github.com/California-Planet-Search/radvel}} \citep{fultonradvel}. The planetary parameters included are the $T_c$, the $P$, and the stellar RV amplitude induced by the planet ($K$). The initial values used for $T_c$ and $P$ were derived from Sect.\ \ref{sec:trchr}. Moreover, as in the photometric case, we included an instrumental offset ($\gamma_{\mathrm{RV}}$).

We employed the criteria established by \cite{trotta2008} based on the Bayesian log-evidence ($\ln \mathcal{Z}$), which was calculated following the method by \citep{diaz16}, to evaluate which is the best RV-only model (results listed in Table\ \ref{tab:logZ}). According to this criterion, when the absolute difference $|\Delta \ln \mathcal{Z}|$\,is greater than five, the model with the higher log evidence is strongly favoured. In cases where $|\Delta \ln \mathcal{Z}|$\,is greater than 2.5 but less than five, the evidence in favour of one model is moderate. If $|\Delta \ln \mathcal{Z}|$\, is greater than one but less than 2.5, the evidence is weak, and when $|\Delta \ln \mathcal{Z}|$\,is less than one, the models are considered indistinguishable. None of the models considering the transiting planet are moderately favoured compared to models without the planet, which seems to indicate that the planet is not detected in the current RV. The models that include only a jitter term are the least strongly favoured. The models composed of two sinusoidal functions are moderately favoured compared to the GP models. 
In addition, the models including a Keplerian signal are indistinguishable or less favoured compared to models without planets, which suggest the absence of any planet 
signal in the RV data. This result is expected since there is no detection of TOI-1135\,b. Therefore, between the activity model with two sinusoidal functions and the activity model with GP, we chose the latter. Although the model with two sinusoidal functions is slightly favoured over the GP model, the planet signal is slightly better detected in the latter. The QP GP models have been widely used in the literature to model the activity of young stars \citep{barr19, klein21, cale21, zicher22, nardiello22, barr22, Mallorquin2023} because the QP variations in activity are better modelled.

\renewcommand{\arraystretch}{1.22} 
\begin{table*}[htbp]
\caption{Model comparison for RV-only analysis of TOI-1135\,b using the difference between Bayesian log-evidences ($\Delta \ln \mathcal{Z}$).}
\begin{center}
\begin{tabular}{cccc|cc|cc}
\hline
\hline
 & & \multicolumn{6}{c}{Activity model}\\
\cline{3-8}
 & & \multicolumn{2}{c}{$\sigma_{\mathrm{jit}}$} & \multicolumn{2}{|c}{2 Sin (P$_{1}$\,$\sim$5.1\,d, P$_{2}$\,$\sim$2.5\,d) + $\sigma_{\mathrm{jit}}$} & \multicolumn{2}{|c}{GP$_\mathrm{GP}$ + $\sigma_{\mathrm{jit}}$}\\
\cline{3-8}
Dataset & Planets & K$^b$[m\,s$^{-1}$] & $\Delta \ln \mathcal{Z} $ & K$^b$[m\,s$^{-1}$] & $\Delta \ln \mathcal{Z}$ & K$^b$[m\,s$^{-1}$] & $\Delta \ln \mathcal{Z}$\\
\hline
CARMENES VIS & 0 & -- & --246.0 & -- & --230.5 & -- & --236.3\\
CARMENES VIS & 1 & 7.7\,$\pm$\,4.6 (21.1) & --247.7 & 5.0\,$\pm$\,3.2 (14.5) & --231.3 & 5.6\,$\pm$\,3.1 (14.9) & --238.6\\
\hline
\end{tabular}
\tablefoot{In the model name, $\sigma_{\mathrm{jit}}$ refers to a jitter term added in quadrature to the RV error bars and "2 Sin" refers to two sinusoidal functions and their periods. All models assume circular orbits, and the amplitudes are given with their 1$\sigma$ uncertainty. The 99.7 percentile of the Keplerian amplitude is provided in parenthesis. The result in bold indicates our adopted RV model.}\\
\label{tab:logZ}
\end{center}
\end{table*}

\subsection{Joint-fit analysis}
\label{sec:joint}

\begin{figure}[ht!]
\includegraphics[width=1\linewidth]{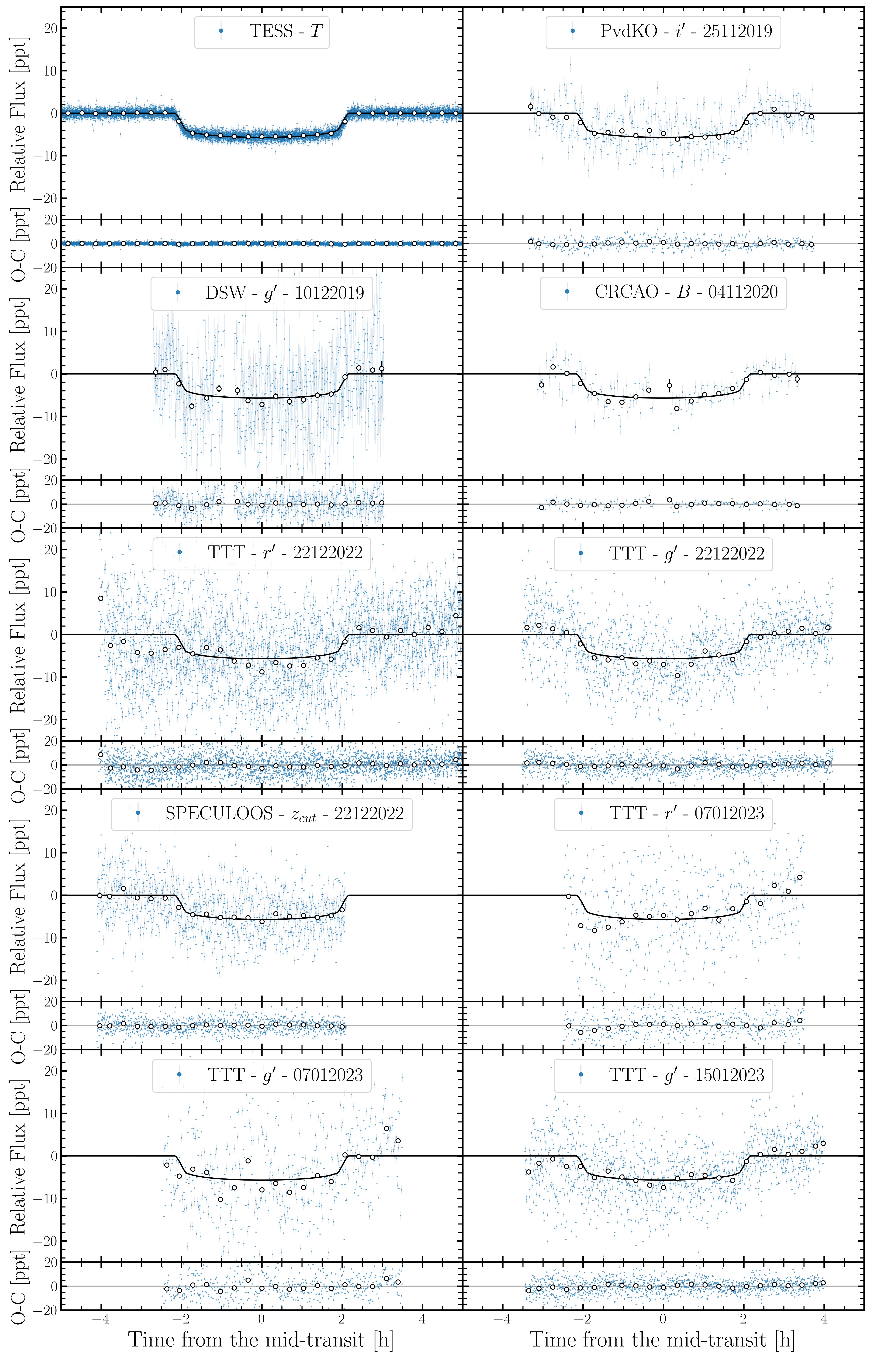}
\caption{Phase-folded light curves of TOI-1135\,b for data of TESS, CRCAO, DSW, PvdKO, SPECULOOS, and TTT. In each sub-panel the photometric data (blue dots) are shown along with the binned data (white dots), the best transit-fit model (black line in the top), and the residuals for the best fit (in the bottom).
\label{fig:PH_folded}}
\end{figure}

\begin{figure}[ht!]
\includegraphics[width=1\linewidth]{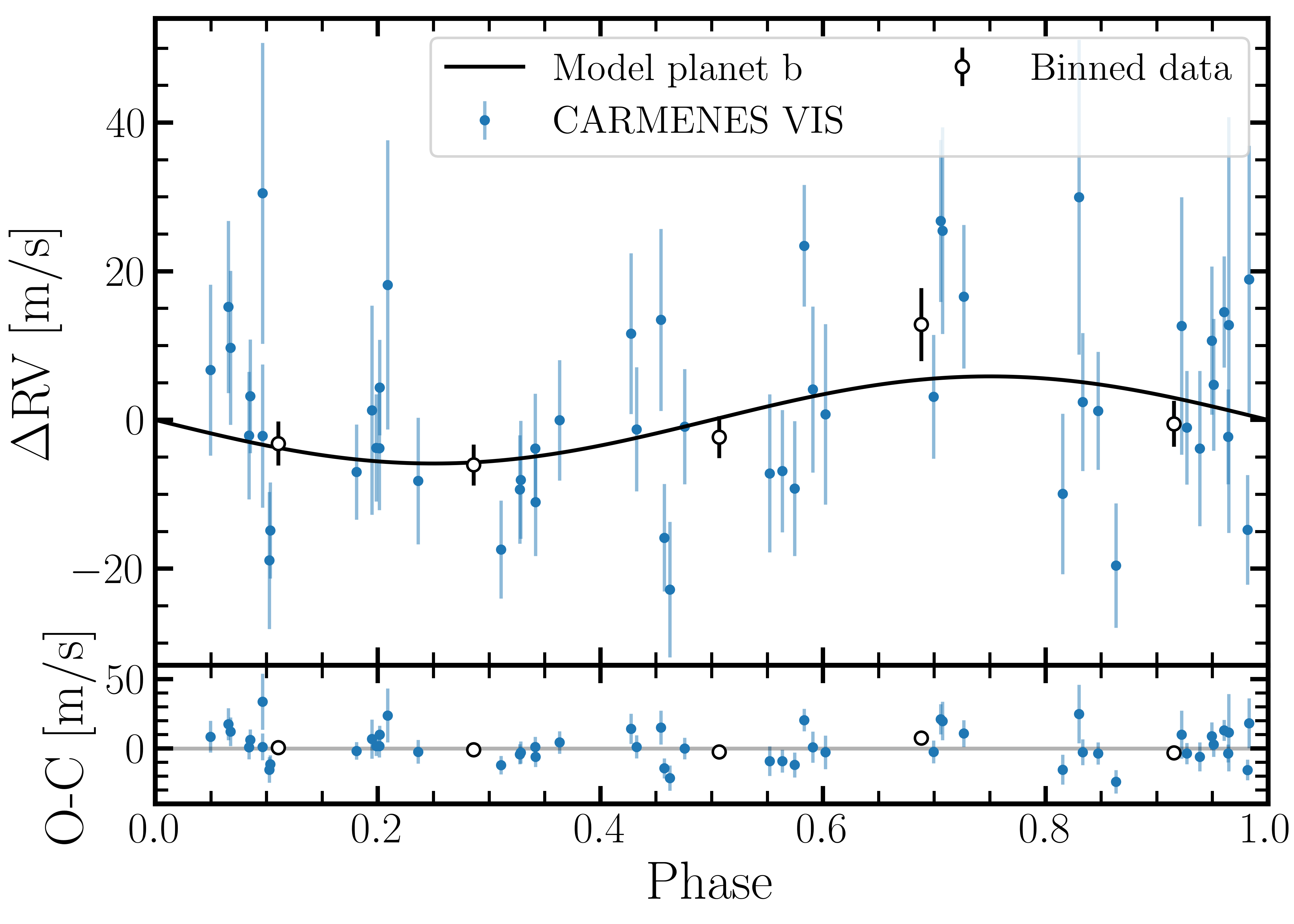}
\caption{Phase-folded RVs for TOI-1135\,b. \textit{Top panel:} CARMENES VIS data (blue dots), binned data (white dots), and the Keplerian model of the joint fit (black line). \textit{Bottom panel:} Residuals for the best fit.}
\label{fig:RV_folded}
\end{figure}

All the data sets were included in a global joint fit combining the transit photometry and RV time series in order to yield more precise parameters for the TOI-1135 system. This global fit includes a transit model over the TESS and the ground-based transit photometry (phase-folded transits in Fig.\ \ref{fig:PH_folded}) and a Keplerian model in the CARMENES RVs (phase-folded RV in Fig.\ \ref{fig:RV_folded}) to obtain the planetary parameters. These planetary parameters included shared parameters such as $T_c$ and $P$ among all the data sets with normal priors, $R_p/R_{\star}$ and $b$ among photometry data sets with uniform priors, and $K$ among RV data sets with uniform priors. We also explored non-circular planetary orbits including uniform priors in $\sqrt{e}\sin{\omega}$ and $\sqrt{e}\cos{\omega}$, which correspond to a re-parametrisation of the eccentricity and the argument of periastron proposed by \cite{anderson2011} and are shared between all the data sets. Furthermore, a stellar activity model generated by GP was used to model the photometric TESS light curves with a dSHO kernel (Sect.\ \ref{sec:trsearch}, Fig.\ \ref{fig:LC_TESS}) and the CARMENES RVs data set with a QP kernel (Sect.\ \ref{sec:rvanalysis}, Fig.\ \ref{fig:RV_curve}). The $\eta_P$ hyperparameter is shared between TESS and the RV data sets, while $\eta_L$ and $\eta_\omega$ are independent among the TESS and RV data sets with uniform priors. However, normal priors were used for covariance amplitude hyperparameters, centred on the standard deviation of each data set. The details of prior and posterior results from the joint fit are listed in Table \ref{tab:joint-fit}. From the posterior results, we derived other planetary parameters of interest, such as the planet radius ($R_p$), orbital inclination ($i$), planet mass ($M_p$), planet bulk density ($\rho_p$), transit depth ($\delta$), and equilibrium temperature ($T_{\mathrm{eq}}$), using error propagation (Table\,\ref{tab:joint-fit_der}).

\begingroup
\begin{table*}
\caption{Prior and posterior parameters of the joint fit for TOI-1135\,b.}\label{tab:joint-fit}
\centering
\begin{tabular}{lccc}
\hline\hline
Parameter & Prior & Posterior ($e$\,$=$\,0) & Posterior ($e$, $\omega$ free)\\
\hline
$T_{c}$[BJD] & $\mathcal{N}$(2459583.2583, 0.1) & 2459583.2582$^{+0.0002}_{-0.0002}$ & 2459583.2581$^{+0.0002}_{-0.0002}$\\
$P$[d] & $\mathcal{N}$(8.0277, 0.1) & 8.027730$^{+0.000003}_{-0.000003}$ & 8.027729$^{+0.000003}_{-0.000003}$\\
$R_{p}/R_{\star}$ & $\mathcal{U}$(0, 0.1) & 0.0697$^{+0.0002}_{-0.0002}$ & 0.0699$^{+0.0003}_{-0.0002}$\\
$b$ & $\mathcal{U}$(0, 1) & 0.007$^{+0.008}_{-0.005}$ & 0.102$^{+0.095}_{-0.071}$\\
$K$[m s$^{-1}$] & $\mathcal{U}$(0, 200) & 5.86$^{+3.11}_{-2.94}$ (15.18) & 6.03$^{+3.09}_{-2.95}$(15.31)\\
$\sqrt{e}\sin\omega$ & $\mathcal{U}$(--1, 1) & ... & --0.144$^{+0.035}_{-0.040}$\\
$\sqrt{e}\cos\omega$ & $\mathcal{U}$(--1, 1) & ... & --0.018$^{+0.319}_{-0.298}$\\
$\gamma_{\mathrm{TESS}}$[ppt] & $\mathcal{U}$(--3$\sigma_{\mathrm{TESS}}$, 3$\sigma_{\mathrm{TESS}}$) & 0.44$^{+0.12}_{-0.12}$ & 0.44$^{+0.12}_{-0.12}$\\
$\gamma_{\mathrm{CRCAO}}$[ppt] & $\mathcal{U}$(--3$\sigma_{\mathrm{CRCAO}}$, 3$\sigma_{\mathrm{CRCAO}}$) & 3.27$^{+0.22}_{-0.22}$ & 3.36$^{+0.22}_{-0.22}$\\
$\gamma_{\mathrm{DSW}}$[ppt] & $\mathcal{U}$(--3$\sigma_{\mathrm{DSW}}$, 3$\sigma_{\mathrm{DSW}}$) & 3.38$^{+0.34}_{-0.34}$ & 3.48$^{+0.34}_{-0.34}$\\
$\gamma_{\mathrm{PvdKO}}$[ppt] & $\mathcal{U}$(--3$\sigma_{\mathrm{PvdKO}}$, 3$\sigma_{\mathrm{PvdKO}}$) & 2.61$^{+0.18}_{-0.18}$ & 2.70$^{+0.19}_{-0.19}$\\
$\gamma_{\mathrm{SPECULOOS}}$[ppt] & $\mathcal{U}$(--3$\sigma_{\mathrm{SPECULOOS}}$, 3$\sigma_{\mathrm{SPECULOOS}}$) & 3.50$^{+0.17}_{-0.17}$ & 3.59$^{+0.17}_{-0.17}$\\
$\gamma_{\mathrm{TTT,} r'\mathrm{, 1}}$[ppt] & $\mathcal{U}$(--3$\sigma_{\mathrm{TTT,} r'\mathrm{, 1}}$, 3$\sigma_{\mathrm{TTT,} r'\mathrm{, 1}}$) & 1.63$^{+0.14}_{-0.14}$ & 1.69$^{+0.14}_{-0.14}$\\
$\gamma_{\mathrm{TTT,} g'\mathrm{, 1}}$[ppt] & $\mathcal{U}$(--3$\sigma_{\mathrm{TTT,} g'\mathrm{, 1}}$, 3$\sigma_{\mathrm{TTT,} g'\mathrm{, 1}}$) & 2.37$^{+0.15}_{-0.15}$ & 2.45$^{+0.15}_{-0.15}$\\
$\gamma_{\mathrm{TTT,} r'\mathrm{, 2}}$[ppt] & $\mathcal{U}$(--3$\sigma_{\mathrm{TTT,} r'\mathrm{, 2}}$, 3$\sigma_{\mathrm{TTT,} r'\mathrm{, 2}}$) & 4.86$^{+0.38}_{-0.38}$ & 4.95$^{+0.38}_{-0.38}$\\
$\gamma_{\mathrm{TTT,} g'\mathrm{, 2}}$[ppt] & $\mathcal{U}$(--3$\sigma_{\mathrm{TTT,} g'\mathrm{, 2}}$, 3$\sigma_{\mathrm{TTT,} g'\mathrm{, 2}}$) & 3.35$^{+0.41}_{-0.41}$ & 3.44$^{+0.41}_{-0.41}$\\
$\gamma_{\mathrm{TTT,} g'\mathrm{, 3}}$[ppt] & $\mathcal{U}$(--3$\sigma_{\mathrm{TTT,} g'\mathrm{, 3}}$, 3$\sigma_{\mathrm{TTT,} g'\mathrm{, 3}}$) & 2.97$^{+0.17}_{-0.17}$ & 3.05$^{+0.17}_{-0.17}$\\
$\gamma_{\mathrm{CARMENES\ VIS}}$[m s$^{-1}$] & $\mathcal{U}$(--3$\sigma_{\mathrm{CARMENES\ VIS}}$, 3$\sigma_{\mathrm{CARMENES\ VIS}}$) & 2.42$^{+11.60}_{-10.91}$ & 1.88$^{+11.52}_{-10.99}$\\
$\sigma_{\mathrm{jit, TESS}}$[ppt] & $\mathcal{U}$(0, 3$\sigma_{\mathrm{TESS}}$) & 0.41$^{+0.01}_{-0.01}$ & 0.40$^{+0.01}_{-0.01}$\\
$\sigma_{\mathrm{jit, CRCAO}}$[ppt] & $\mathcal{U}$(0, 3$\sigma_{\mathrm{CRCAO}}$) & 2.36$^{+0.20}_{-0.19}$ & 2.37$^{+0.20}_{-0.19}$\\
$\sigma_{\mathrm{jit, DSW}}$[ppt] & $\mathcal{U}$(0, 3$\sigma_{\mathrm{DSW}}$) & 7.68$^{+0.29}_{-0.27}$ & 7.66$^{+0.28}_{-0.28}$\\
$\sigma_{\mathrm{jit, PvdKO}}$[ppt] & $\mathcal{U}$(0, 3$\sigma_{\mathrm{PvdKO}}$) & 3.50$^{+0.15}_{-0.14}$ & 3.49$^{+0.15}_{-0.14}$\\
$\sigma_{\mathrm{jit, SPECULOOS}}$[ppt] & $\mathcal{U}$(0, 3$\sigma_{\mathrm{SPECULOOS}}$) & 5.34$^{+0.13}_{-0.12}$ & 5.32$^{+0.12}_{-0.12}$\\
$\sigma_{\mathrm{jit, TTT,} r'\mathrm{, 1}}$[ppt] & $\mathcal{U}$(0, 3$\sigma_{\mathrm{TTT,} r'\mathrm{, 1}}$) & 8.36$^{+0.10}_{-0.10}$ & 8.35$^{+0.10}_{-0.10}$\\
$\sigma_{\mathrm{jit, TTT,} g'\mathrm{, 1}}$[ppt] & $\mathcal{U}$(0, 3$\sigma_{\mathrm{TTT,} g'\mathrm{, 1}}$) & 5.75$^{+0.11}_{-0.11}$ & 5.75$^{+0.11}_{-0.10}$\\
$\sigma_{\mathrm{jit, TTT,} r'\mathrm{, 2}}$[ppt] & $\mathcal{U}$(0, 3$\sigma_{\mathrm{TTT,} r'\mathrm{, 2}}$) & 8.84$^{+0.28}_{-0.27}$ & 8.82$^{+0.27}_{-0.26}$\\
$\sigma_{\mathrm{jit, TTT,} g'\mathrm{, 2}}$[ppt] & $\mathcal{U}$(0, 3$\sigma_{\mathrm{TTT,} g'\mathrm{, 2}}$) & 9.41$^{+0.30}_{-0.28}$ & 9.41$^{+0.30}_{-0.29}$\\
$\sigma_{\mathrm{jit, TTT,} g'\mathrm{, 3}}$[ppt] & $\mathcal{U}$(0, 3$\sigma_{\mathrm{TTT,} g'\mathrm{, 3}}$) & 5.97$^{+0.12}_{-0.12}$ & 5.95$^{+0.12}_{-0.12}$\\
$\sigma_{\mathrm{jit, CARMENES\ VIS}}$[m s$^{-1}$] & $\mathcal{U}$(0, 3$\sigma_{\mathrm{CARMENES\ VIS}}$) & 8.61$^{+3.05}_{-3.62}$ & 8.63$^{+3.03}_{-3.69}$\\
$q_{1, T}$ & $\mathcal{N}$(0.28, 0.01) & 0.26$^{+0.01}_{-0.01}$ & 0.27$^{+0.01}_{-0.01}$\\
$q_{2, T}$ & $\mathcal{N}$(0.36, 0.01) & 0.35$^{+0.01}_{-0.01}$ & 0.36$^{+0.01}_{-0.01}$\\
$q_{1, B}$ & $\mathcal{N}$(0.65, 0.01) & 0.65$^{+0.01}_{-0.01}$ & 0.65$^{+0.01}_{-0.01}$\\
$q_{2, B}$ & $\mathcal{N}$(0.41, 0.01) & 0.40$^{+0.01}_{-0.01}$ & 0.40$^{+0.01}_{-0.01}$\\
$q_{1, g'}$ & $\mathcal{N}$(0.60, 0.01) & 0.60$^{+0.01}_{-0.01}$ & 0.60$^{+0.01}_{-0.01}$\\
$q_{2, g'}$ & $\mathcal{N}$(0.40, 0.01) & 0.40$^{+0.01}_{-0.01}$ & 0.40$^{+0.01}_{-0.01}$\\
$q_{1, i'}$ & $\mathcal{N}$(0.29, 0.01) & 0.29$^{+0.01}_{-0.01}$ & 0.29$^{+0.01}_{-0.01}$\\
$q_{2, i'}$ & $\mathcal{N}$(0.36, 0.01) & 0.36$^{+0.01}_{-0.01}$ & 0.36$^{+0.01}_{-0.01}$\\
$q_{1, z_{cut}}$ & $\mathcal{N}$(0.23, 0.01) & 0.23$^{+0.01}_{-0.01}$ & 0.23$^{+0.08}_{-0.01}$\\
$q_{2, z_{cut}}$ & $\mathcal{N}$(0.35, 0.01) & 0.35$^{+0.01}_{-0.01}$ & 0.35$^{+0.10}_{-0.01}$\\
$q_{1, r'}$ & $\mathcal{N}$(0.39, 0.01) & 0.39$^{+0.01}_{-0.01}$ & 0.39$^{+0.01}_{-0.01}$\\
$q_{2, r'}$ & $\mathcal{N}$(0.37, 0.01) & 0.37$^{+0.01}_{-0.01}$ & 0.37$^{+0.01}_{-0.01}$\\
$\eta_{\sigma_1, \mathrm{TESS}}$ & $\mathcal{N}$($\sigma_{\mathrm{TESS}}$, 0.8) & 3.57$^{+0.59}_{-0.49}$ & 3.57$^{+0.58}_{-0.49}$\\
$\eta_{\sigma_2, \mathrm{TESS}}$ & $\mathcal{N}$($\sigma_{\mathrm{TESS}}$, 0.8) & 2.00$^{+0.18}_{-0.14}$ & 1.99$^{+0.18}_{-0.14}$\\
$\eta_{\sigma, \mathrm{CARMENES\ VIS}}$ & $\mathcal{N}$($\sigma_{\mathrm{CARMENES\ VIS}}$, 5) & 23.44$^{+4.37}_{-4.02}$ & 23.45$^{+4.29}_{-4.02}$\\
$\eta_{L_1, \mathrm{TESS}}$ & $\mathcal{U}$($P_\mathrm{rot}$, 2500) & 33.21$^{+19.55}_{-11.83}$ & 32.64$^{+18.99}_{-11.43}$\\
$\eta_{L_2, \mathrm{TESS}}$ & $\mathcal{U}$($P_\mathrm{rot}$, 2500) & 5.64$^{+1.02}_{-0.41}$ & 5.64$^{+1.03}_{-0.41}$\\
$\eta_{L, \mathrm{RV}}$ & $\mathcal{U}$($P_\mathrm{rot}$, 150) & 70.21$^{+52.77}_{-33.91}$ & 69.66$^{+52.61}_{-33.60}$\\
$\eta_{P_{\mathrm{rot}}}$ & $\mathcal{U}$(3.5, 10) & 5.10$^{+0.03}_{-0.02}$ & 5.09$^{+0.03}_{-0.02}$\\
$\eta_{\omega, \mathrm{RV}}$ & $\mathcal{U}$(0.1, 1.0) & 0.32$^{+0.11}_{-0.10}$ & 0.32$^{+0.12}_{-0.10}$\\
\hline
$\Delta \ln \mathcal{Z}$ & ... & --118219.6 & --118101.4\\
\hline
\end{tabular}
\tablefoot{The prior label of $\mathcal{N}$ and $\mathcal{U}$ represent normal and uniform distributions, respectively. The 99.7 percentile of the planet mass and bulk density are provided in parenthesis.}\\
\end{table*}
\endgroup

\begingroup
\renewcommand{\arraystretch}{1.25} 
\begin{table}
\caption{Derived parameters of the joint fit for TOI-1135\,b.}\label{tab:joint-fit_der}
\centering
\begin{tabular}{l@{\hskip 0.01in}c@{\hskip 0.04in}c}
\hline\hline
Parameter & Posterior ($e$\,$=$\,0) & Posterior ($e$, $\omega$ free)\\
\hline
$a/R_{\star}$ & 15.12$^{+0.40}_{-0.40}$ & 15.12$^{+0.40}_{-0.40}$\\
$a$ [AU] & 0.082$^{+0.003}_{-0.003}$ & 0.082$^{+0.003}_{-0.003}$\\
$R_{p}$ [R$_{\mathrm{Jup}}$] & 0.805$^{+0.020}_{-0.020}$ & 0.807$^{+0.020}_{-0.020}$\\
$R_{p}$ [R$_{\oplus}$] & 9.020$^{+0.227}_{-0.227}$ & 9.042$^{+0.229}_{-0.228}$\\
$i$ [$^\circ$] & 89.97$^{+0.03}_{-0.02}$ & 89.62$^{+0.35}_{-0.27}$\\
$e$ & ... & 0.07$^{+0.10}_{-0.04}$\\
$\omega$ [rad] & ... & --1.67$^{+1.28}_{-1.11}$\\
$M_p$ [M$_{\mathrm{Jup}}$] & 0.062$^{+0.033}_{-0.031}$ (0.162) & 0.064$^{+0.033}_{-0.031}$ (0.163)\\
$M_p$ [M$_{\oplus}$] & 19.84$^{+10.53}_{-9.96}$ (51.42) & 20.43$^{+10.49}_{-10.00}$ (51.89)\\
$\rho_p$ [g\,cm$^{-3}$] & 0.16$^{+0.09}_{-0.08}$ (0.41) & 0.16$^{+0.08}_{-0.08}$ (0.42)\\
$t_{12}$ [h] & 0.28$^{+0.01}_{-0.01}$ & 0.29$^{+0.01}_{-0.01}$\\
$t_{14}$ [h] & 4.05$^{+0.11}_{-0.11}$ & 4.12$^{+0.11}_{-0.11}$\\
$\delta$ [ppt] & 4.86$^{+0.03}_{-0.03}$ & 4.88$^{+0.04}_{-0.03}$\\
$T_{\mathrm{eq}}$(A\,=\,0) [K] & 1198.9$^{+21.3}_{-21.3}$ & 1198.9$^{+21.3}_{-21.3}$\\
$T_{\mathrm{eq}}$(A\,=\,0.6) [K] & 953.4$^{+17.0}_{-17.0}$ & 953.4$^{+17.0}_{-17.0}$\\
$u_{1,\mathrm{T}}$ & 0.36$^{+0.01}_{-0.01}$ & 0.37$^{+0.01}_{-0.01}$\\
$u_{2,\mathrm{T}}$ & 0.15$^{+0.01}_{-0.01}$ & 0.15$^{+0.01}_{-0.01}$\\
$u_{1,\mathrm{B}}$ & 0.65$^{+0.01}_{-0.01}$ & 0.66$^{+0.01}_{-0.01}$\\
$u_{2,\mathrm{B}}$ & 0.15$^{+0.01}_{-0.01}$ & 0.15$^{+0.01}_{-0.01}$\\
$u_{1,\mathrm{g'}}$ & 0.61$^{+0.01}_{-0.01}$ & 0.61$^{+0.01}_{-0.01}$\\
$u_{2,\mathrm{g'}}$ & 0.16$^{+0.01}_{-0.01}$ & 0.16$^{+0.01}_{-0.01}$\\
$u_{1,\mathrm{i'}}$ & 0.39$^{+0.01}_{-0.01}$ & 0.39$^{+0.01}_{-0.01}$\\
$u_{2,\mathrm{i'}}$ & 0.15$^{+0.01}_{-0.01}$ & 0.15$^{+0.01}_{-0.01}$\\
$u_{1,\mathrm{z'}}$ & 0.34$^{+0.01}_{-0.01}$ & 0.34$^{+0.01}_{-0.01}$\\
$u_{2,\mathrm{z'}}$ & 0.14$^{+0.01}_{-0.01}$ & 0.14$^{+0.01}_{-0.01}$\\
$u_{1,\mathrm{r'}}$ & 0.46$^{+0.01}_{-0.01}$ & 0.46$^{+0.01}_{-0.01}$\\
$u_{2,\mathrm{r'}}$ & 0.16$^{+0.01}_{-0.01}$ & 0.16$^{+0.01}_{-0.01}$\\
\hline
\end{tabular}
\tablefoot{The 99.7 percentile of the planet mass and bulk density are provided in parenthesis.}\\
\end{table}
\endgroup

\subsection{Transit timing variations}

\begin{figure}[ht!]
\includegraphics[width=1\linewidth]{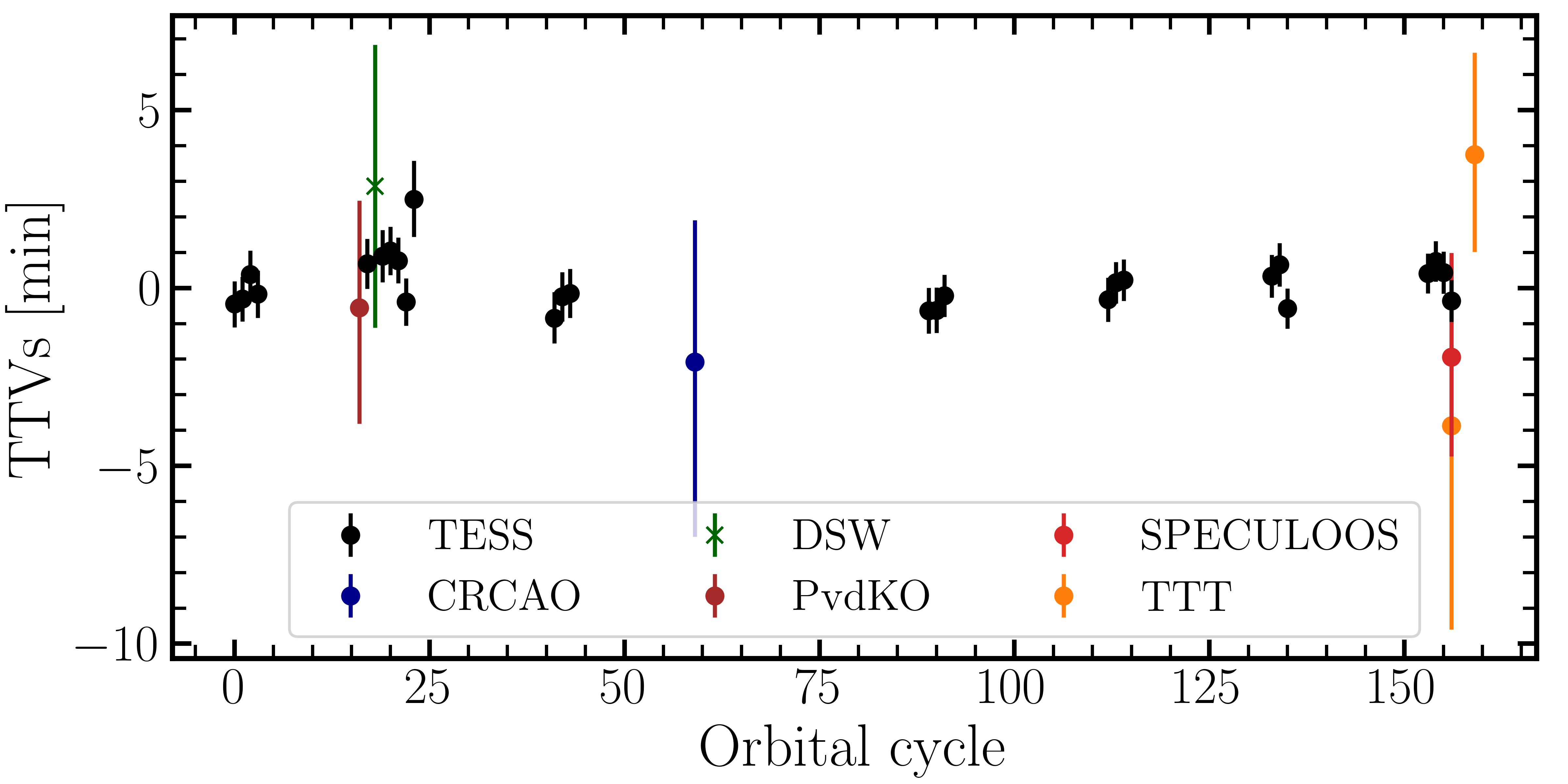}
\caption{Transit timing variations for the planet TOI-1135\,b. The TTVs are represented as coloured dots with their 1$\sigma$ uncertainty. The transits with error bars larger than 10 min have been removed for clarity.
\label{fig:TTV}}
\end{figure}

Transit timing variations can be used to estimate the planetary masses of a system or, at least, to indicate the existence of additional companions in the system. Thanks to the TESS and ground-based photometry, we had a temporal coverage spanning $\sim$1300 days with a total of 35 transits. We built an O-C diagram fitting individual transit times ($T_{c}$) for each transit (Fig.\ \ref{fig:TTV}) from the posteriors obtained in the joint-fit analysis (Sect.\ \ref{sec:joint}). The TTVs obtained with ground-based photometry have large uncertainties. However, the TESS data have an rms of 0.7 min, with a median error bar of 0.6 min. Consequently, no significant TTVs were detected in the system that would allow us to constrain additional candidates.

\section{Validation and false positive scenarios}
\label{sec:fap}

Several astrophysical scenarios, such as diluted eclipsing binaries or grazing transits, can mimic planetary transit signals. The following analysis rules out most of these false positive scenarios, confirming the bona fide planetary nature of TOI-1135\,b.

\begin{itemize}
\item \textbf{The transit signal is a result of instrumental artefacts.}

The detection of the transits from ground-based photometry (see Sect.\ \ref{sec:trchr}) discards the possibility of instrumental false positives associated with the TESS satellite system. Furthermore, the period of these transits does not align with any known periodicities in the TESS satellite system, such as momentum dumps.

\item \textbf{The transit signal is a result of stellar variability.}

The observed stellar variability is much larger (peak-to-peak amplitude of $\sim$20 ppt, Fig.\,\ref{fig:LC_TESS}) than the depth of the transits ($\sim$4.9 ppt, Table\,\ref{tab:joint-fit_der}), and the orbital period of TOI-1135\,b is not a multiple of the stellar rotation period inferred from the TESS light curves. Moreover, we expected a lower stellar activity at redder wavelengths, but the transit depths do not show any chromaticity (Fig.\ \ref{fig:trchr}).

\item \textbf{The transit signal is a result of a blended source.}

Given that the observed transit has a well-defined and flat shape (Fig.\ \ref{fig:PH_folded}), we could rule out grazing transits where a source partially eclipses the star. In addition, our spectra do not show blended spectral lines nor RV variations large enough to indicate a stellar-mass companion ($<$100 m\,s$^{-1}$, Fig.\ \ref{fig:RV_curve}). Due to the larger TESS pixel scale, it is common for the source to be contaminated by nearby stars in crowded fields, as in the case of TOI-1135, with a star at 40\arcsec (Fig.\ \ref{fig:TPF}, star number 2). However, the ground-based observations with apertures of less than 10\arcsec confirm that the transit is not occurring on star number 2, and as the transits have a similar transit depth (Fig.\ \ref{fig:trchr}), the dilution factor is negligible. 

The NESSI speckle image shows a star 0.973\arcsec away with a brightness 5.39 mag fainter than TOI-1135 at 832 nm (Fig.\ \ref{fig:NESSI}). However, the SAI image rules out other sources within 1\arcsec with magnitudes 7.2 times fainter at 625 nm (Fig.\ \ref{fig:SAI}). This could indicate that the star observed in NESSI is significantly redder than TOI-1135 ($R-I \gtrsim$\,1.8 mag). If the star is assumed to be bound, the limit on the $R-I$ colour would indicate that the star is a late-type M dwarf ($\gtrsim$\,M5-M6V), but the measured magnitude difference at 832 nm suggests that the star is closer to an earlier M dwarf ($\sim$M3V). This discrepancy likely means that the companion is a foreground source unrelated to the target. From $Gaia$ DR3 observations, we obtained a RUWE value of 0.974, which indicates that the astrometric solution is consistent with a single star model. However, the IPDfmp parameter, which provides the fraction of windows for which the image parameter determination algorithm has identified multiple peaks, is 3\%, which is a number close to zero but not zero and is consistent with the star observed in the NESSI image. To check if the observed signal is transiting TOI-1135 or on the nearby companion, we used the \cite{Vanderburg2019} formula:

\begin{equation}
\Delta m  \lesssim 2.5\log_{10} \left[ \left(\frac{t_{12}}{t_{13}} \right)^2 \frac{1}{\delta}\right], \\
\label{eq:van}
\end{equation}

which, using the transit ingress time ($t_{12}$), the time between the first and third contact ($t_{13}$), and the transit depth ($\delta$; parameters derived in Table\ \ref{tab:joint-fit_der}), gives a lower limit for the faintest companion that could cause this transit. We obtained $\Delta m_{TESS}$\,$\lesssim$\,0.4 mag at a 3$\sigma$ level of confidence, meaning that a source of similar brightness to TOI-1135 is needed to reproduce the observed transit. Additionally, if the companion is clearly redder than TOI-1135, we would expect to see some chromaticity in the transit depth, which is not the case (Fig.\ \ref{fig:trchr}). In a more quantitative way, we can calculate the expected transit depth at 832 nm if a complete eclipse occurs in the star that is 5.39 mag fainter than TOI-1135 (Sect.\ \ref{sec:nessi}). By eclipsing 100\% of the star, we obtained a transit depth of 7.0 ppt, which is not compatible at 832 nm according to Fig\,\ref{fig:trchr}. It would be necessary to eclipse the star by $\sim$120\% to reach 8.3 ppt. Similarly, if we assume a star at 625 nm at the detection limit of 7.2 mag (Sect.\ \ref{sec:sai}), we obtain a transit depth of 1.3 ppt, making it necessary to eclipse the star by more than 600\% to reach the observed 8.5 ppt in Fig\,\ref{fig:trchr}. Therefore, the transit is not occurring on the fainter companion. Finally, \cite{Hord2023} carried out a vetting process statistically validating TOI-1135\,b using the \texttt{vespa} \citep{Morton2012, Morton2015} and \texttt{TRICERATOPS} \citep{Giancalone2020, Giancalone2021} software packages, calculating the false positive probability and classifying TOI-1135\,b as a validated planet.

\end{itemize}

\section{Discussion on planet properties}
\label{sec:disc}

\subsection{Planet characterisation}

Based on the posterior parameters obtained from the joint analysis (Sect.\ \ref{sec:joint}), we determined a planetary radius of $R_p^b$=\,9.02\,$\pm$\,0.23\,R$_\oplus$ ($R_p^b$=\,0.84\,$\pm$\,0.02 R$_{\rm{Jup}}$). Furthermore, our best fit model for a planet provides a mass result of 19.8$^{+10.5}_{-10.0}$ M$_\oplus$, a $\sim$2$\sigma$ detection. Thus, we set an upper limit of $M_p^b$\,$<$\,51.4\,M$_\oplus$ ($M_p^b$\,$<$\,0.16 M$_{\rm{Jup}}$) at a 3$\sigma$ confidence level. These results imply an upper limit for the bulk density of 0.41 g\,cm$^{-3}$. Although, according to Bayesian log-evidence, the model with a non-circular orbit is moderately favoured over the solution with a circular orbit. This model indicates that if the planet eccentricity is non-zero, it is very low (e\,$<$\,0.26 at 2$\sigma$). Since we do not have a robust detection of the planet in our RV data, we adopt the solution with a circular orbit that has a smaller number of free parameters. Considering a range of planetary albedos ($A_{\rm Bond}$) from 0.6 to 0.0, we estimate an equilibrium temperature ($T_\mathrm{eq}$) in the range of 950--1200\,K for TOI-1135\,b.

\subsection{Mass loss rate}
\label{sec:massloss}

The expected mass loss rate in a planet atmosphere can be estimated if we assume an energy-limited approach and an atmosphere dominated by hydrogen. We need to know the stellar XUV (X-ray+EUV, 5--920\,\AA{}) irradiation at the planet's orbit, which is normally calculated either using a coronal model or a scaling law based on the X-ray stellar luminosity. However, no X-ray flux measurements are available to date for TOI-1135\@. To overcome this limitation, we made use of the relation between the stellar rotation period and the Rossby number and the X-ray stellar luminosity \citep{Wright2011}. Using the parameters as listed in Table\ \ref{tab:stellar_parameters} and the magnitudes $V$ and $K$, we calculated $\log L_{\rm X}$\,=\,29.48 for TOI-1135\@. We made use of the scaling law calculated in \citet{Sanz2022} to estimate the EUV (100--920\,\AA{}) stellar luminosity of $\log L_{\rm EUV,H}$\,=\,29.90\@. We then calculated the energy-limited mass loss rate of TOI-1135~b using Eq.~7 of \citet{Sanz2011}, of 7.4$\times10^{12}$\,g\,s$^{-1}$, or 39\,M$_\oplus$\,Gyr$^{-1}$.

The calculated mass loss rate indicates that although this rate will decrease with time, the planet may lose most of its atmosphere in a few hundred million years. TOI-1135\,b is likely an interesting candidate to search for atmospheric photoevaporation. The \ion{He}{i}~10830 triplet is sensitive to atmospheric photoevaporation, and the line formation in the planet atmosphere is strongly influenced by the stellar He-ionising irradiation in the XUV range ($<$504\,\AA{}). We calculated the EUV flux in the 100--504\,\AA{} range using the relations in \citet{Sanz2022}, $\log L_{\rm EUV,He}$\,=\,29.62\@. The XUV flux at the planet's orbit is $\sim$\,43000\,erg\,s$^{-1}$\,cm${-2}$, which is slightly larger than the flux reaching the orbit of WASP-69\,b \citep{Nortmann2018}, where the triplet was clearly detected. Thus, TOI-1135\,b is a firm candidate to search for signatures of photoevaporation in the \ion{He}{i}~10830 triplet. The study of its atmosphere and search for an extended \ion{He}{I} atmosphere will be presented in a future paper (Orell-Miquel et al., in prep.)

\subsection{Mass-insolation-radius diagram}

\begin{figure*}[ht!]
\includegraphics[width=1\linewidth]{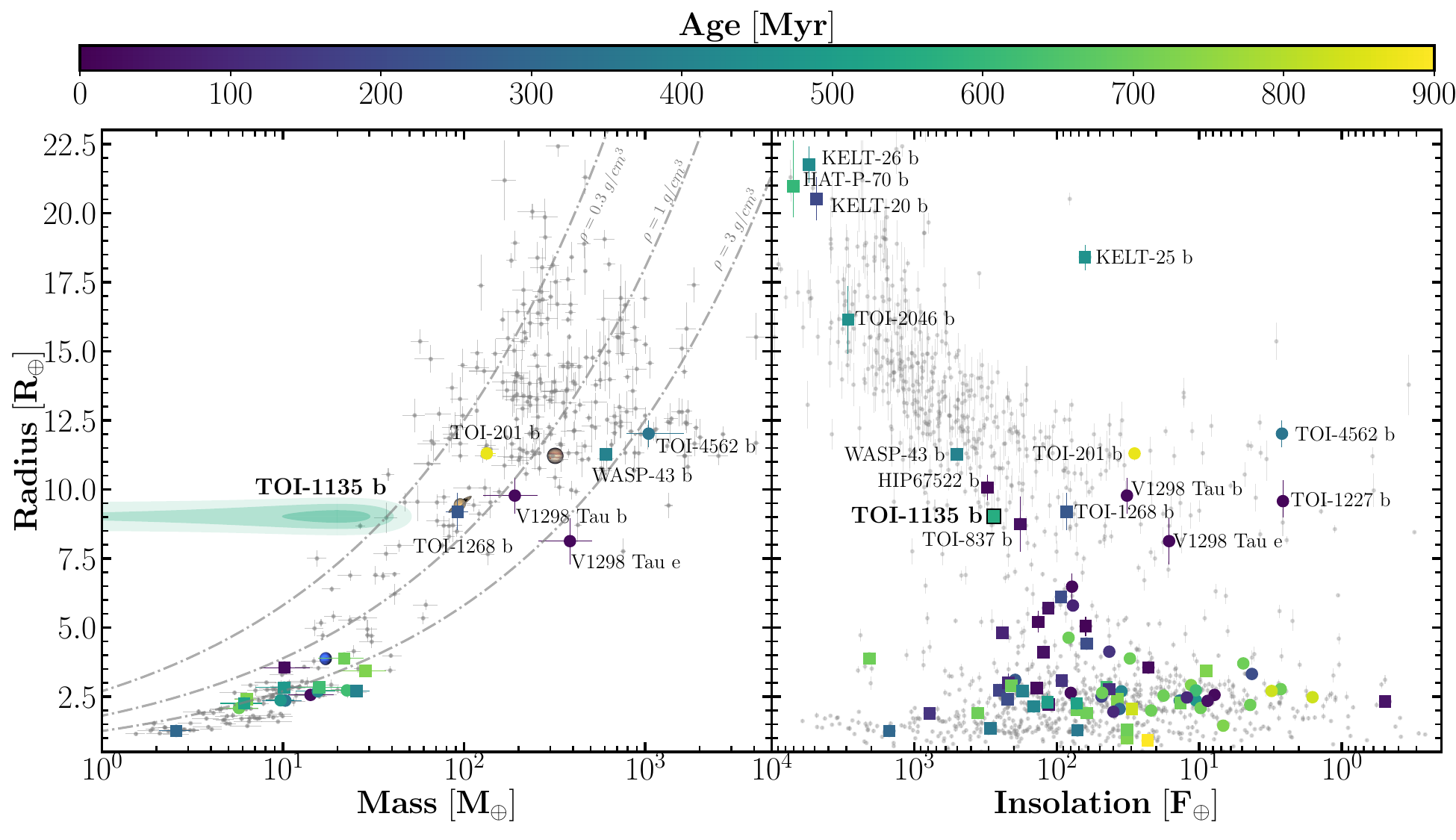}
\caption{Mass-insolation-radius diagram for TOI-1135\,b, together with all known exoplanets (grey dots) with a precision better than 8\% in radius (through transit) and 20\% in mass (from RV) from the Extrasolar Planets Encyclopaedia. The population of young transiting exoplanets ($\leq$ 900 Myr) with measured masses is plotted as coloured dots, according to their ages. The uncertainties on TOI-1135\,b are shown as coloured shaded regions with 1, 2, and 3$\sigma$ levels of confidence. In the left panel, the iso-density lines are displayed as dashed grey lines. Saturn, Jupiter, and Neptune are also depicted for reference.
\label{fig:RM}}
\end{figure*}

In Fig.\ \ref{fig:RM}, we present a mass-insolation-radius diagram of known transiting exoplanets (gray dots) from the Extrasolar Planets Encyclopaedia,\footnote{\url{http://exoplanet.eu/}} with radius uncertainties better than 8\% determined through the transit method and mass uncertainties better than 20\% from the RV method. Additionally, we have overplotted in colour known exoplanets younger than 900 Myr from our own collection. The dots and squares indicate young exoplanets with orbital periods lower and higher than 10 days, respectively, while the colour indicates the age. In the left panel of Fig.\ \ref{fig:RM}, the green-shaded regions represent the radius and mass posterior distribution for TOI-1135\,b with 1, 2, and 3$\sigma$ significance intervals. We did not include the population of planets whose masses were estimated by TTVs due to several studies, including \citet{Hadden2017}, have shown that the population of planets whose masses were estimated by TTVs are less dense than the population of planets for which the masses were estimated through RV. In the right panel of the figure, its radius and period are shown as square error bars with its 1$\sigma$ uncertainty.

The left panel in Fig.\ \ref{fig:RM} shows a clear overdensity of exoplanets in the upper part ($R_p$\,$>$\,8--10\,R$_\oplus$) that corresponds to giant gas planets, while in the lower part the overdensity of planets corresponds to "small" planets ($R_p$\,$<$\,4--5\,R$_\oplus$). Both sequences are separated by a region with a lower frequency of planets. In the mass-radius diagram, TOI-1135\,b is located in the lower-left limit of giant gas planets or Jupiter-type planets, suggesting mostly a gaseous composition of H and He. In addition, only two young giant planets have mass measurements with an orbital period below 10 days, namely, TOI-1268\,b \citep[$M_p$\,=\,102\,$\pm$\,11\,M$_\oplus$, P$_{\rm{orb}}$=8.158 days;][]{Subjak2022} and WASP-43\,b \citep[$M_p$\,=\,635\,$\pm$\,25\,M$_\oplus$, P$_{\rm{orb}}$=0.813 days;][]{Davoudi2021}. Both of these planets seem to fit with the sequence of field planets and with the planets of our Solar System, as TOI-1268\,b and WASP-43\,b are comparable with Saturn and Jupiter, respectively. However, the mass of TOI-1135\,b could also be compatible with a planet with the mass of Neptune or even less, indicating that it could be a puffed-up Neptune-type planet with different characteristics than TOI-1268\,b and WASP-63\,b.

In the insolation-radius panel (right-hand side panel in Fig.\ \ref{fig:RM}), TOI-1135\,b is in the upper limit of the transition zone between giant gaseous planets and Neptune-like planets. While KELT-26\,b, HAT-P-70\,b, KELT-20\,b, and TOI-2046\,b clearly belong to the hot Jupiter group, the cases of WASP-43\,b, HIP67522\,b, TOI-837\,b, TOI-1268\,b, or TOI-1135\,b are not clear. If we assume a maximum mass of 50\,M$_\oplus$ for TOI-1135\,b and apply the loss of mass rate of its atmosphere calculated in Sect.\ \ref{sec:massloss}, we observe that in a few hundred million years, TOI-1135\,b will lose most of its atmosphere, decreasing its radius and mass and aligning with the group of small planets such as Neptune. The only planets with similar characteristics of measured mass are TOI-1268\,b \citep[100--380 Myr;][]{Subjak2022} and WASP-43\,b \citep[$\sim$400 Myr;][]{Davoudi2021}, whose masses are significantly larger, hence allowing them to retain their atmospheres and belong to the group of gaseous giants. On the other hand, measuring the masses of HIP67522\,b \citep[$\sim$17 Myr;][]{Rizzuto2020} and TOI-837\,b \citep[$\sim$35 Myr;][]{Bouma2020} is key to understanding whether these planets are Saturn-like or Neptune-mass planets with an extended atmosphere that they could eventually lose, implying a subsequent evolution shifting the planet to the lower-left region of the mass-radius diagram.

Formation models predict that gaseous giant planets, such as TOI-1135\,b, formed beyond the ice line and then migrated inwards in timescales of less than 10 Myr \citep{Williams2011}. This, together with the loss of mass of 39\,M$_{\oplus}$\,Gyr$^{-1}$, allows us to estimate that TOI-1135\,b could have originally had a mass equal to or less than 90\,M$_\oplus$, so giant planets with masses less than Saturn are susceptible to losing all or most of their atmosphere in their early stages if they receive enough radiation from a host star, which might be the case of TOI-1135\,b.

\section{Conclusions}
\label{sec:concl}

Our study presents the discovery and characterisation of TOI-1135\,b, a Saturn-size planet transiting a young solar-type star. We inferred the stellar parameters and estimated an age between 125--1000 Myr from different activity-age indicators. We also simultaneously fitted the TESS light curves, the ground-based transit photometry, and the CARMENES RV data in order to derive the planetary parameters. We confirm the planetary nature of TOI-1135\,b, and we derived an orbital period of 8.027730\,$\pm$\,0.000003 days for TOI-1135\,b, measured a radius of 9.02\,$\pm$\,0.23 R$_\oplus$, and set an upper limit on the mass of $<$\,51.4 M$_\oplus$ at a 3$\sigma$ confidence level. 

We conclude that the internal composition of TOI-1135\,b is mostly gaseous. Although it has a radius the size of Saturn, its mass could be compatible with Neptune, indicating an extended atmosphere most probably due to strong stellar radiation. The large mass loss rate of TOI-1135\,b indicates that the planet will lose most of its atmosphere in a few hundred million years. The mass determination and atmospheric study of TOI-1135\,b and other younger giant planets will be key to understanding the formation and atmospheric evolution of planets smaller than Saturn and explaining the lower frequency of planets with sizes between Neptune and Saturn.

\begin{acknowledgements}
MM, NL, and VJSB acknowledges support from the Agencia Estatal de Investigaci\'on del Ministerio de Ciencia e Innovaci\'on (AEI-MCINN) under grant PID2019-109522GB-C53\@.
This paper includes data collected by the $TESS$ mission. Funding for the $TESS$ mission is provided by the NASA Explorer Program. We acknowledge the use of public TOI Release data from pipelines at the $TESS$ Science Office and at the $TESS$ Science Processing Operations Center. Resources supporting this work were provided by the NASA High-End Computing (HEC) Program through the NASA Advanced Supercomputing (NAS) Division at Ames Research Center for the production of the SPOC data products. This research has made use of the Exoplanet Follow-up Observation Program website, which is operated by the California Institute of Technology, under contract with the National Aeronautics and Space Administration under the Exoplanet Exploration Program.
This work has made use of data from the European Space Agency (ESA) mission {\it Gaia} (\url{https://www.cosmos.esa.int/gaia}), processed by the {\it Gaia} Data Processing and Analysis Consortium (DPAC, \url{https://www.cosmos.esa.int/web/gaia/dpac/consortium}). Funding for the DPAC has been provided by national institutions, in particular the institutions participating in the {\it Gaia} Multilateral Agreement.
CARMENES is an instrument at the Centro Astron\'omico Hispano en Andaluc\'ia (CAHA) at Calar Alto (Almer\'{\i}a, Spain), operated jointly by the Junta de Andaluc\'ia and the Instituto de Astrof\'isica de Andaluc\'ia (CSIC). CARMENES was funded by the Max-Planck-Gesellschaft (MPG), the Consejo Superior de Investigaciones Cient\'{\i}ficas (CSIC), the Ministerio de Econom\'ia y Competitividad (MINECO) and the European Regional Development Fund (ERDF) through projects FICTS-2011-02, ICTS-2017-07-CAHA-4, and CAHA16-CE-3978, and the members of the CARMENES Consortium (Max-Planck-Institut f\"ur Astronomie, Instituto de Astrof\'{\i}sica de Andaluc\'{\i}a, Landessternwarte K\"onigstuhl, Institut de Ci\`encies de l'Espai, Institut f\"ur Astrophysik G\"ottingen, Universidad Complutense de Madrid, Th\"uringer Landessternwarte Tautenburg, Instituto de Astrof\'{\i}sica de Canarias, Hamburger Sternwarte, Centro de Astrobiolog\'{\i}a and Centro Astron\'omico Hispano-Alem\'an), with additional contributions by the MINECO, the Deutsche Forschungsgemeinschaft through the Major Research Instrumentation Programme and Research Unit FOR2544 ''Blue Planets around Red Stars'', the Klaus Tschira Stiftung, the states of Baden-W\"urttemberg and Niedersachsen, and by the Junta de Andaluc\'{\i}a.
Some of the bbservations in the paper made use of the NN-EXPLORE Exoplanet and Stellar Speckle Imager (NESSI). NESSI was funded by the NASA Exoplanet Exploration Program and the NASA Ames Research Center. NESSI was built at the Ames Research Center by Steve B. Howell, Nic Scott, Elliott P. Horch, and Emmett Quigley.
J.d.W. and MIT gratefully acknowledge financial support from the Heising-Simons Foundation, Dr. and Mrs. Colin Masson and Dr. Peter A. Gilman for Artemis, the first telescope of the SPECULOOS network situated in Tenerife, Spain. The ULiege's contribution to SPECULOOS has received funding from the European Research Council under the European Union's Seventh Framework Programme (FP/2007-2013) (grant Agreement n$^\circ$ 336480/SPECULOOS), from the Balzan Prize and Francqui Foundations, from the Belgian Scientific Research Foundation (F.R.S.-FNRS; grant n$^\circ$ T.0109.20), from the University of Liege, and from the ARC grant for Concerted Research Actions financed by the Wallonia-Brussels Federation. M.G. is F.R.S.-FNRS Research Director'.
L.G.M.\ and P.A.R.\ acknowledge support from NSF grant no.\ 1952545.
The postdoctoral fellowship of KB is funded by F.R.S.-FNRS grant T.0109.20 and by the Francqui Foundation.
This article includes observations made at the Two-Meter Twin Telescope (TTT) that Light Bridges operates in the Teide Observatory of the Instituto de Astrof\'isica de Canarias (IAC, Tenerife, Spain). The Observing Time Rights (DTO) used for this research have been funded by IAC.

\end{acknowledgements}

\bibliographystyle{aa.bst} 
\bibliography{biblio.bib}

\begin{appendix}

\section{Radial velocity data}

\onecolumn
\begin{longtable}{crr}
\caption{RV data from CARMENES VIS.} \label{tab:CARMV_RV}\\ 
\hline\hline
Time & \multicolumn{1}{c}{RV} & \multicolumn{1}{c}{$\sigma$} \\

[BJD] & [m\,s$^{-1}$] & [m\,s$^{-1}$] \\ 
\hline
\endfirsthead
\caption{continued.} \\ 
\hline
Time & \multicolumn{1}{c}{RV} & \multicolumn{1}{c}{$\sigma$} \\

[BJD] & [m\,s$^{-1}$] & [m\,s$^{-1}$] \\ 
\hline
\endhead
\hline
\endfoot
\hline\hline
\endlastfoot
2460006.3790 & 35.77 & 13.89\\ 
2460007.5006 & --16.33 & 7.95\\ 
2460012.3759 & --3.23 & 12.24\\ 
2460014.5590 & 46.87 & 9.66\\ 
2460015.3914 & 18.53 & 21.17\\ 
2460015.6567 & --47.27 & 8.36\\ 
2460017.5293 & 11.85 & 20.24\\ 
2460018.3498 & --19.02 & 7.20\\ 
2460018.6504 & --14.03 & 8.51\\ 
2460019.3924 & 20.69 & 7.92\\ 
2460019.6711 & 29.07 & 8.09\\ 
2460032.5227 & --12.12 & 6.37\\ 
2460032.6619 & --30.89 & 7.38\\ 
2460033.3370 & --6.38 & 11.58\\ 
2460033.6327 & --34.56 & 9.22\\ 
2460034.3737 & 20.50 & 14.07\\ 
2460034.4860 & 41.73 & 19.46\\ 
2460037.4218 & --7.16 & 9.06\\ 
2460037.6459 & --10.21 & 12.13\\ 
2460038.4763 & 5.32 & 10.88\\ 
2460039.3565 & 4.58 & 10.77\\ 
2460039.5013 & 23.43 & 9.28\\ 
2460040.4316 & 22.81 & 9.96\\ 
2460040.7013 & 18.40 & 17.98\\ 
2460041.3808 & --13.57 & 10.36\\ 
2460041.6684 & --21.87 & 6.46\\ 
2460042.4554 & 2.82 & 8.32\\ 
2460043.3321 & --41.87 & 6.58\\ 
2460043.5805 & --32.62 & 7.29\\ 
2460044.6568 & 22.36 & 7.76\\ 
2460057.5692 & 15.96 & 8.58\\ 
2460057.6671 & 11.65 & 9.64\\ 
2460059.5226 & --1.91 & 7.28\\ 
2460059.6341 & 9.26 & 7.38\\ 
2460060.3661 & 17.46 & 8.37\\ 
2460060.6057 & --11.09 & 9.10\\ 
2460061.4180 & --21.36 & 8.23\\ 
2460061.6363 & --13.90 & 11.18\\ 
2460064.4286 & --5.43 & 10.44\\ 
2460064.6368 & 22.33 & 27.95\\ 
2460066.3732 & --15.57 & 6.40\\ 
2460066.5390 & --9.42 & 6.42\\ 
2460068.3532 & --1.62 & 10.82\\ 
2460068.5929 & --37.66 & 7.24\\ 
2460069.3531 & --16.35 & 10.64\\ 
2460069.6030 & 27.32 & 8.17\\ 
2460070.5362 & 19.23 & 8.31\\ 
2460072.3630 & 9.14 & 7.64\\ 
2460072.6310 & 37.14 & 7.48\\ 
2460073.3480 & 0.00 & 11.50\\ 
2460073.6349 & --16.35 & 7.65\\ 
2460080.3539 & 35.12 & 17.33\\ 
2460080.5841 & 22.70 & 8.87\\ 
\hline
\end{longtable}

\end{appendix}

\end{document}